\documentclass[aps,pra,reprint,showpacs]{revtex4-1}

\usepackage{graphicx}

\usepackage{amssymb}
\usepackage{amsmath}

\usepackage{bm}

\newcommand{\rbr}[1]{\ensuremath{\left(#1\right)}}
\newcommand{\sbr}[1]{\ensuremath{\left[#1\right]}}

\newcommand{\set}[1]{\left\{#1\right\}}

\newcommand{\diff}[2]{\ensuremath{\frac{d#1}{d#2}}}

\newcommand{\pdiff}[2]{\ensuremath{\frac{\partial #1}{\partial #2}}}
\newcommand{\pdoublediff}[2]{\ensuremath{\frac{\partial^2 #1}{\partial #2 ^2}}}
\newcommand{\pmixdiff}[3]{\ensuremath{\frac{\partial^2 #1}{\partial #2 \partial #3}}}

\newcommand{\mbf}[1]{\mathbf{#1}}
\newcommand{\mbb}[1]{\mathbb{#1}}
\newcommand{\wt}[1]{\widetilde{#1}}

\newcommand{\abs}[1]{\left|#1\right|}
\newcommand{\ME}[3]{\ensuremath{\left\langle\left. #1\right. \right| #2 \left| \left. #3 \right. \right\rangle}}

\usepackage[dvipsnames]{color}

\begin{document}

%\preprint{}

\title{Metric space analysis of systems immersed in a magnetic field} %Title of paper



\author{P. M. Sharp}
\email[]{pms510@york.ac.uk}
%\thanks{}
\affiliation{Department of Physics and York Centre for Quantum Technologies, University of York, York, YO10 5DD, United Kingdom}

\author{I. D'Amico}
\email[]{irene.damico@york.ac.uk}
%\thanks{}
\affiliation{Department of Physics and York Centre for Quantum Technologies, University of York, York, YO10 5DD, United Kingdom}


\date{\today}

\begin{abstract}

Understanding the behavior of quantum systems subject to magnetic fields is of fundamental importance and underpins quantum technologies.
However, modeling these systems is a complex task, because of many-body interactions and because many-body approaches such as density functional
theory get complicated by the presence of a vector potential into the system Hamiltonian. We use the metric space approach to quantum mechanics to study the effects of varying the magnetic vector potential on
quantum systems. The application of this technique to model systems in the ground state provides insight into the fundamental mapping at the core of current density functional theory,
which relates the many-body wavefunction, particle density and paramagnetic current density. We show that the role of the paramagnetic current density in this relationship becomes crucial
when considering states with different magnetic quantum numbers, $m$. Additionally, varying the magnetic field uncovers a richer complexity for the ``band structure'' present in ground
state metric spaces, as compared to previous studies varying scalar potentials. The robust nature of the metric space approach is strengthened by demonstrating the gauge invariance of
the related metric for the paramagnetic current density. We go beyond ground state properties and apply this approach to excited states. The results suggest that, under specific conditions,
a universal behavior may exist for the relationships between the physical quantities defining the system.

\end{abstract}


\maketitle %\maketitle must follow title, authors, abstract and \pacs


% If in two-column mode, this environment will change to single-column format so that long equations can be displayed. 
%\begin{widetext}
%$$\mbox{put long equation here}$$
%\end{widetext}

\section{Introduction}

Systems immersed in magnetic fields are a fundamental research topic, as is the case, for example, for atoms immersed in strong fields \cite{Thirumalai2009,Thirumalai2014}, and
are also an integral part of emerging quantum technologies, such as quantum computation, which utilise quantum systems controlled
or otherwise affected by magnetic fields. For example the inhomogeneous magnetic field generated by the nuclei's spins decreases quantum
coherence of electron spin qubits in III-V quantum dots \cite{Mehring2003},
while full polarization of the spin bath through an applied magnetic field suppresses electron-spin decoherence in nitrogen-vacancy centers and nitrogen impurities in diamond \cite{Takahashi2008}.
Understanding systems immersed in a magnetic field at a quantum level is therefore of both fundamental and technological importance.

However, the presence of a magnetic field introduces additional complexity to the system's Hamiltonian. In fact, as opposed to a confining potential, which is defined by a scalar potential,
$V\rbr{\mbf{r}}$, the magnetic field is defined by $\mbf{B}\rbr{\mbf{r}}=\nabla\times\mbf{A}\rbr{\mbf{r}}$, where $\mbf{A}\rbr{\mbf{r}}$ is a {\it vector} potential.
To account for its presence, density functional theory (DFT) must be extended to current density functional theory (CDFT).
In this paper we will use the metric space approach to quantum mechanics
\cite{D'Amico2011,Arthacho2011,D'Amico2011b,Sharp2014} to study the effect on quantum systems of changing the vector potential and we will carefully consider the implications of
the results for CDFT. This is particularly relevant as there are still open questions with respect to the fundamentals of this theory (see, for example, Refs.
\cite{Taut2009,Tellgren2012,Lieb2013,Tellgren2014,Laestadius2015}).

The metric space approach involves the derivation of ``natural'' metrics from conservation laws to assign
a distance between two physically relevant functions \cite{Sharp2014}. In recent work these metrics were applied to the basic variables
of both standard DFT \cite{D'Amico2011,Arthacho2011,D'Amico2011b,Nagy2011} and CDFT \cite{Sharp2014}.
In Refs.~\cite{D'Amico2011} and \cite{Sharp2014} it was demonstrated that the core theorems of DFT and CDFT,
respectively, indeed represent mappings between metric spaces: This helps understanding the power of the metric space approach and why its use has already allowed for the discovery of additional properties of these core theorems.
The results in Ref.~\cite{Sharp2014}, pertaining to systems with fixed magnetic fields, considered only the effects of varying the scalar potential; in this paper we demonstrate how the metric space approach to quantum mechanics
can be applied to analyse systems while varying the vector potential and hence the magnetic field.

A significant issue for any theory involving magnetic fields is gauge transformations of the scalar and vector potentials. The magnetic field,
along with all physical observables, is gauge invariant. Hence, in order to properly describe the distances between physical quantities, the
metrics we derive must be robust against gauge transformations. However, quantities such as the wavefunction and the paramagnetic current density
are gauge variant and changes in the vector potential can result in gauge transformations for these quantities. Thus, here we will extend 
the metric space approach to ensure that the metrics associated with these quantities are gauge invariant.

We will provide further insight into the fundamental mappings between key physical quantities at the core of CDFT by studying the ground state of model systems as the vector potential is
varied. In particular we will examine how the ``band structure'' introduced into ground-state metric spaces by the presence of a magnetic field \cite{Sharp2014} responds to changes in the field.
To complement this picture, we will apply the metric space approach to quantum mechanics to explore the properties of excited states.
Results will also help with validating the conclusions from the ground-state analysis.

The rest of this paper is organised as follows. In Sec.~\ref{metric} we briefly review how functions obeying integral conservation laws can be cast as
metric spaces and the application of this approach to the wavefunction, particle density, and paramagnetic current density. Section~\ref{gauge}
demonstrates how the gauge properties of the paramagnetic current density are accounted for when forming the related metric space in order to ensure that the metric is
gauge invariant. In Sec.~\ref{structures} we examine how the ``band structure'' present in metric spaces for ground states is affected by variations in the
magnetic field and the relevance of this for CDFT. Section~\ref{excite} goes beyond ground states. In Sec.~\ref{conclusion} we present a summary and our conclusions.
We use atomic units $\hbar=e=m_e=1/4\pi\epsilon_{0}=1$ throughout this paper.

\section{Metric Spaces for Physical Functions} \label{metric}

In Ref.~\cite{Sharp2014} a general procedure for deriving metric spaces from conservation laws was presented. For completeness in this section we
briefly review this procedure and the properties of these metric spaces.

A metric space $\rbr{X,D}$ consists of a non-empty set of points $X$ and a metric $D: X\times X \rightarrow \mbb{R}$, which assigns a distance between any
two elements of $X$. For all $a,b,c \in X$ the metric must satisfy the following axioms \cite{Sutherland2009,Megginson1998}:
\begin{align}
D\rbr{a,b} &\geqslant 0\ \text{and}\ D\rbr{a,b}=0 \iff a=b, \label{axiom1}\\
D\rbr{a,b} &= D\rbr{b,a}, \label{axiom2}\\
D\rbr{a,b} &\leqslant D\rbr{a,c}+D\rbr{c,b}. \label{axiom3}
\end{align}
These axioms are known as positivity, symmetry, and the triangle inequality respectively.

\begin{figure}[t]
\begin{center}
 \includegraphics[width=\columnwidth]{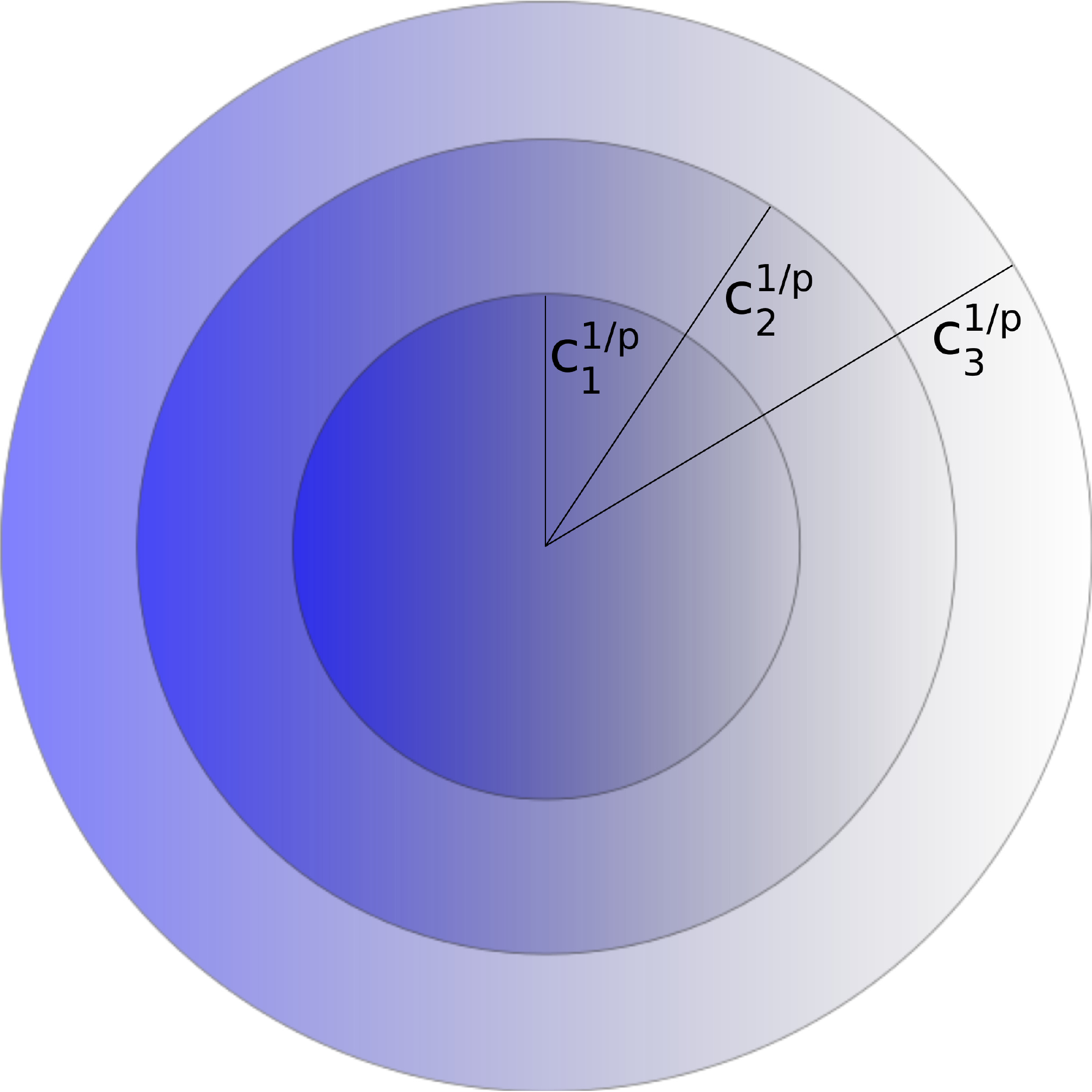}
 \caption{(Color online) Sketch of the structure of the ``onion-shell'' geometry, consisting of a series of concentric spheres. The first three spheres are shown, with radii
 $c_{i}^{\frac{1}{p}}, i=1,2,3$.}
 \label{spheres}
\end{center}
\end{figure}

Consider now a conservation law of the form
\begin{equation} \label{conservation}
 \int\abs{f\rbr{\mbf{x}}}^{p} d\mbf{x} = c,
\end{equation}
with $c$ a finite, positive constant and $\mbf{x}=(x_1,\ldots,x_N)$ including any spatial or spin coordinate in any dimensionality. Then for each $p$ such
that $1\leqslant p<\infty$, the entire set of functions that satisfies Eq.~(\ref{conservation}) forms an $L^{p}$ vector space. Then the corresponding metric
\begin{equation} \label{Lp_metric}
 D_{f}\rbr{f_{1},f_{2}}=\sbr{\int\abs{f_1\rbr{\mbf{x}}-f_2\rbr{\mbf{x}}}^p d\mbf{x}}^{\frac{1}{p}},
\end{equation}
also applies to the \emph{restricted} set of \emph{physical} functions obeying the conservation law~(\ref{conservation})~\cite{Sharp2014}. As $c$ spans the set of its physically allowed values
$\set{c_{i}}$, the metric (\ref{Lp_metric}) imposes on its metric space an ``onion-shell'' geometry that consists of a series of concentric spheres with radii
$c_{i}^{\frac{1}{p}}$, as sketched in Fig.~\ref{spheres}.

We note that the procedure developed in Ref.~\cite{Sharp2014} can be extended to conservation laws of the form
\begin{equation} \label{sum_law}
 \sum_{i=1}^{n}\abs{f_{i}}^{p}=c,
\end{equation}
as the $l^{p}$ vector spaces for sums are directly analogous to the $L^{p}$ spaces for integrals \cite{Megginson1998}. In this case the induced metric will be
\begin{equation} \label{lp_metric}
 D_{f}\rbr{f_{1},f_{2}}=\sbr{\sum_{i=1}^{n}\abs{f_{1_{i}}-f_{2_{i}}}^{p}}^{\frac{1}{p}}.
\end{equation}
Thus, we have a general procedure to construct a metric for any conservation law that is, or can be cast, in the form of
Eq.~(\ref{conservation}) or (\ref{sum_law}). We can therefore state that such conservation laws induce metrics on the set of related physical
functions. As they descend directly from conservation laws, these ``natural'' metrics are non-trivial and contain the relevant physics.

Following this procedure the metrics
\begin{align}
 D_{\psi}(\psi_{1},\psi_{2})=&\left[\int\rbr{\abs{\psi_{1}}^{2}+\abs{\psi_{2}}^{2}}d\mbf{r}_{1} \ldots d\mbf{r}_{N}\right.\nonumber\\
 &\left.-2\abs{\int\psi_{1}^{*}\psi_{2}d\mbf{r}_{1} \ldots d\mbf{r}_{N}}\right]^{\frac{1}{2}}, \label{dpsi}\\
 D_{\rho}(\rho_{1},\rho_{2})=&\int\abs{\rho_{1}(\mbf{r})-\rho_{2}(\mbf{r})} d\mbf{r}, \label{drho}\\
 D_{\mbf{j}_{p_{\perp}}}\rbr{\mbf{j}_{p_{1}},\mbf{j}_{p_{2}}}=&\int\abs{\set{\mbf{r}\times\sbr{\mbf{j}_{p_{1}}\rbr{\mbf{r}}-\mbf{j}_{p_{2}}\rbr{\mbf{r}}}}_{z}} d\mbf{r}, \label{dj}
\end{align}
for wavefunctions $\psi$ \cite{D'Amico2011},
particle densities $\rho$ \cite{D'Amico2011}, and paramagnetic current densities $\mbf{j}_{p}=\rbr{j_{p_{x}},j_{p_{y}},j_{p_{z}}}$ \cite{Sharp2014}
have been introduced. These metrics follow from, respectively, the conservation of wavefunction norm, of particle number, and of the $z$ component
of the angular momentum; the latter can be expressed as Eq.~(\ref{conservation}), when using the relation \cite{Sharp2014}
\begin{equation} \label{Lz}
 \int\sbr{\mbf{r}\times\mbf{j}_{p}\rbr{\mbf{r}}}_{z} d\mbf{r}=\ME{\psi}{\hat{L}_{z}}{\psi}=m.
\end{equation}
For paramagnetic current densities, Eq.~(\ref{Lz}) directly imposes an equivalence relation on the set of all
paramagnetic current densities because $\sbr{\mbf{r}\times\mbf{j}_{p}\rbr{\mbf{r}}}_{z}$ is independent of $j_{p_{z}}$. As a result, $D_{\mbf{j}_{p_{\perp}}}$
is a metric defined on a set of equivalence classes of paramagnetic current densities, with the classes characterised by paramagnetic current
densities with the same transverse component $\mbf{j}_{p_{\perp}}=\rbr{j_{p_{x}},j_{p_{y}}}$.

The radii of the concentric spheres in the ``onion-shell'' geometry of the aforementioned metric spaces are: $\sqrt{N}$ for the wavefunction metric space,
\footnote{We follow the same convention as Ref.~\cite{D'Amico2011}, where wavefunctions are normalised to the particle number $N$} $N$ for the particle
density metric space, and $\abs{m}$ for the paramagnetic current density metric space \cite{D'Amico2011,Sharp2014}. When considering individual spheres,
the diameter of the sphere imposes an upper bound on the value of the distance.

\section{Gauge Invariance of Metrics} \label{gauge}

When dealing with electromagnetic fields, it is important to consider the choice of gauge.
The scalar and vector potentials in the Hamiltonian are not unique, as a change of gauge transforms the potentials according to
\begin{equation} \label{gauge_trans}
 V'\rbr{\mbf{r}}=V\rbr{\mbf{r}}+c', \quad \mbf{A}'\rbr{\mbf{r}}=\mbf{A}\rbr{\mbf{r}}-\nabla\chi,
\end{equation}
where $c'$ is a constant and $\chi\rbr{\mbf{r}}$ is a scalar field \cite{Vignale1987}. These transformations preserve the electromagnetic fields and
all physical observables.

With regard to the quantities we consider in this paper, the particle density is gauge invariant, but the wavefunction and paramagnetic current
density are not. After a change of gauge, the wavefunction undergoes a unitary transformation, which is given by \cite{Cohen1977}
\begin{equation} \label{psi_trans}
 \psi'\rbr{\mbf{r}}=e^{\sbr{{i\chi\rbr{\mbf{r}}}}}\psi\rbr{\mbf{r}}.
\end{equation}
The paramagnetic current density transforms according to \cite{Vignale1987}
\begin{equation} \label{jp_trans}
 \mbf{j}'_{p}\rbr{\mbf{r}}=\mbf{j}_{p}\rbr{\mbf{r}}+\rho\rbr{\mbf{r}}\nabla\chi.
\end{equation}
Thus, when considering changes in the vector potential, we must be aware of the effect of gauge transformations on the physical quantities we are considering.
Our metrics are constructed to provide non-trivial information that is physically relevant; they are based in fact on conservation laws.
It is paramount then that they are also gauge invariant. The issue of ensuring that the wavefunction metric is gauge invariant has been discussed in Ref.~\cite{D'Amico2011};
we provide a formal review of the approach in Appendix~\ref{wave_metric}. In this paper we wish to discuss gauge invariance with respect to the paramagnetic current density metric.

\subsection{Gauge invariance for the paramagnetic current density metric}

To consider the gauge properties of $D_{\mbf{j}_{p_{\perp}}}(\mbf{j}_{p_{1}},\mbf{j}_{p_{2}})$, first of all we require that $\mbf{j}_{p_{1}}\rbr{\mbf{r}}$
and $\mbf{j}_{p_{2}}\rbr{\mbf{r}}$ are within the same gauge. Then, applying the gauge transformation~(\ref{jp_trans}), we obtain
\begin{align}
 D_{\mbf{j}_{p_{\perp}}}\bigl(\mbf{j}'_{p_{1}},\mbf{j}'_{p_{2}}\bigr)=&\int\abs{\set{\mbf{r}\times\sbr{\mbf{j}'_{p_{1}}\rbr{\mbf{r}}-\mbf{j}'_{p_{2}}\rbr{\mbf{r}}}}_{z}} d\mbf{r}, \nonumber\\
 =&\int\abs{\rbr{\mbf{r}\times\set{\mbf{j}_{p_{1}}\rbr{\mbf{r}}-\mbf{j}_{p_{2}}\rbr{\mbf{r}}\right.\right.\right. \nonumber\\
 &\left.\left.\left.+\sbr{\rho_{1}\rbr{\mbf{r}}-\rho_{2}\rbr{\mbf{r}}}\nabla\chi}}_{z}} d\mbf{r}.\label{dj_trans} 
\end{align}
Equation~(\ref{dj_trans}) states that, in general, the paramagnetic current density distance
defined by Eq.~(\ref{dj}) is modified by a gauge transformation.
This seems to contradict the fact that we base our metrics on conservation laws, which must be gauge invariant.
In order to reconcile this apparent contradiction let us explore more closely which quantities are gauge variant and which are the ones that must be conserved.

With reference to Eq.~(\ref{Lz}), the measurable physical quantity that must be conserved by gauge transformations is $m$, which, in the gauge
chosen, corresponds to the component $\hat{L}_z$ of the angular momentum. However, it is crucial to note that $\hat{L}_z$ is not (nor need be) gauge invariant.

In fact the operator $\hat{L}_{z}$ is defined as
\begin{equation}
 \hat{L}_{z}=\sum_{i=1}^{N}\sbr{\mbf{r}_{i}\times\hat{\mbf{p}}_{i}}_{z},
\end{equation}
where $\hat{\mbf{p}}$ is the canonical linear momentum
$\hat{\mbf{p}}=-i\nabla$. Although $\mbf{r}$ is gauge invariant, $\hat{\mbf{p}}$ is gauge variant and therefore so is $\hat{L}_{z}$.
In the following we wish to extend Eq.~(\ref{dj}) so that the metric associated with the conservation of $m$ is indeed gauge invariant.

We consider a system for which there exists at least one gauge such that $[\hat{L}_{z},\hat{H}]=0$, with $\hat{H}$ the system Hamiltonian.
We name this the reference gauge and refer to its vector potential as $\mbf{A}_{ref}\rbr{\mbf{r}}$ and to its paramagnetic current density as $\mbf{j}_{p_{ref}}\rbr{\mbf{r}}$.
In this reference gauge the set $\set{m}$ corresponds to the eigenvalues of $\hat{L}_{z}$ and both equalities in the relation~(\ref{Lz}) hold. The set $\set{m}$ is then
a constant of motion and in this gauge it represents the $z$ component of the angular momentum.

We now focus on the generic gauge corresponding to a generic vector potential $\mbf{A}\rbr{\mbf{r}}$. 
In this generic gauge, the first equality of Eq.~(\ref{Lz}) holds, but the second equality holds only if $\hat{L}_{z}$ is a constant of motion in this gauge.
Here we consider the quantity
\begin{equation} \label{tilde_jp}
 \wt{\mbf{j}}_p\rbr{\mbf{r}}\equiv\mbf{j}_p\rbr{\mbf{r}}-\rho\rbr{\mbf{r}}\nabla\chi_{ref}\\
\end{equation}
and the operator
\begin{equation} \label{tilde_L}
 \wt{L}_{z}\equiv\sum_{i=1}^{N}\sbr{\mbf{r}\times\rbr{{\mbf{\hat{p}}}-\nabla\chi_{ref}}}_{z},
\end{equation}
where $\nabla\chi_{ref}$ is defined by $\mbf{A}=\mbf{A}_{ref}-\nabla\chi_{ref}$.
We note that $\wt{\mbf{j}}_p\rbr{\mbf{r}}$ is gauge invariant, as, from Eq.~(\ref{jp_trans}),
\begin{equation}
\wt{\mbf{j}}_p\rbr{\mbf{r}}\equiv\mbf{j}_{p_{ref}}\rbr{\mbf{r}}
\end{equation}
always. It follows that
\begin{equation} \label{tilde_jp_m}
 \int\sbr{\mbf{r}\times\wt{\mbf{j}}_{p}\rbr{\mbf{r}}}_{z} d\mbf{r}=m
\end{equation}
independently of the gauge. Furthermore, by using the definition~(\ref{tilde_jp}), where $\rho\rbr{\mbf{r}}=\int\abs{\psi\rbr{\mbf{r}\mbf{r}_{2}\ldots\mbf{r}_{N}}}^2 d\mbf{r}_{2}\ldots d\mbf{r}_{N}$,
and the first equality of Eq.~(\ref{Lz}), which holds regardless of whether or not $\hat{L}_{z}$ is a constant of motion, we obtain
\begin{align}
\int\sbr{\mbf{r}\times\wt{\mbf{j}}_{p}\rbr{\mbf{r}}}_{z} d\mbf{r}&=
\ME{\psi}{\hat{L}_{z}}{\psi}-\int\sbr{\mbf{r}\times\rho\rbr{\mbf{r}}\nabla\chi_{ref}}_{z}d\mbf{r},\nonumber\\
&=\ME{\psi}{\hat{L}_{z}}{\psi}-\ME{\psi}{\rbr{\mbf{r}\times\nabla\chi_{ref}}_{z}}{\psi},\nonumber\\
&=\ME{\psi}{\wt{{L}}_{z}}{\psi}.\label{demo}
\end{align}
This demonstrates that Eq.~(\ref{tilde_L}) defines the operator
associated to the conservation law~(\ref{tilde_jp_m}) independently of the gauge. In particular, comparison of Eqs.~(\ref{tilde_jp_m}) and~(\ref{demo}) shows
that indeed $\wt{{L}}_{z}$ is the operator whose eigenvalues are $\set{m}$ independently of the gauge.
\footnote{We note that $\tilde{L}_{z}$ is related to the gauge invariant $z$ component of the moment of mechanical momentum
$\hat{K}_{z}=\hat{L}_{z}+\sbr{\mathbf{r}\times\mathbf{A}\rbr{\mathbf{r}}}_{z}$ as
$\tilde{L}_{z}=\hat{K}_{z}-\sbr{\mathbf{r}\times\mathbf{A}_{ref}\rbr{\mbf{r}}}_{z}$, but that $\hat{K}_{z}$ would not be a constant
of motion in all gauges, that is, its eigenvalues are generally different from $\set{m}$. Likewise $\tilde{\mathbf{j}}_{p}$
does not coincide with the gauge invariant total current density $\mathbf{j}\rbr{\mathbf{r}}=\mathbf{j}_{p}\rbr{\mathbf{r}}+\rho\rbr{\mbf{r}}\mathbf{A}\rbr{\mathbf{r}}$.}.

Here $\wt{{L}}_{z}$ reduces to ${\hat{L}}_{z}$ in the reference gauge and in all gauges where ${\hat{L}}_{z}$ is a constant of motion, as should be expected.
This is because the limited set of gauges for which $[{\hat{L}}_{z},H]=0$ holds, is the same within which both ${\hat{L}}_{z}$ and $\sbr{\mbf{r}\times{\mbf{j}}_{p}\rbr{\mbf{r}}}_{z}$
are unaffected by gauge transformations. These gauges correspond to vector potentials of the form
\begin{equation} \label{gauges}
 \mbf{A}\rbr{\mbf{r}}=\sbr{x\alpha+y\beta,y\alpha-x\beta,\gamma},
\end{equation}
where $\alpha,\beta,$ and $\gamma$ are all arbitrary functions of $\rbr{x^{2}+y^{2},z}$. These vector potentials are linked by gauge transformations of
the form $\chi\rbr{x^{2}+y^{2},z}$. Demonstration of these statements is quite long and the details are given in Appendix~\ref{vector_potentials}.
Using the conservation law~(\ref{tilde_jp_m}), we derive the metric,
\begin{equation} \label{dj_tilde}
  D_{\wt{\mbf{j}}_{p_{\perp}}}\rbr{\wt{\mbf{j}}_{p_{1}},\wt{\mbf{j}}_{p_{2}}}=\int\abs{\set{\mbf{r}\times\sbr{\wt{\mbf{j}}_{p_{1}}\rbr{\mbf{r}}-\wt{\mbf{j}}_{p_{2}}\rbr{\mbf{r}}}}_{z}} d\mbf{r},
\end{equation}
for the gauge invariant current density, $\wt{\mbf{j}}_{p}$.

\section{Metric analysis of ground states for varying magnetic fields} \label{structures}

In the presence of a magnetic field the metric spaces for ground-state wavefunctions, particle densities and
paramagnetic current densities are characterised by a ``band structure'' \cite{Sharp2014}.
This is significant as identification and characterisation of ground-state properties is very important in several contexts but far from obvious.
This ``band structure'' originates from the conservation law involving the paramagnetic current density $\mbf{j}_{p}$. In $\mbf{j}_{p}$ metric space, when considering variations in the scalar
potential \cite{Sharp2014}, the ``band structure'' is formed by
spherical segments of allowed and forbidden distances on the concentric spheres, at least for the systems analysed.
The specific arc length of these segments varies depending on the radius $\abs{m}$ of the sphere.
Here we wish to investigate how this ``band structure'' responds to changes in the magnetic field.

\subsection{Model systems}

We focus on two atomiclike model systems with uniform, time-independent magnetic fields $\mbf{B}=\omega_{c}c\mbf{\hat{z}}$ applied, where $c$ is the speed of light. Both systems
consist of two electrons in harmonic confinement but with different electron-electron interactions. One system, known as the magnetic Hooke's Atom (HA),
has a two-body Coulomb interaction \cite{Taut1994,Taut2009}, whereas in the other electrons interact via an inverse square interaction (ISI), the relative
strength of which can be varied through an interaction parameter $\alpha$ \cite{Quiroga1993}. The Hamiltonians for these systems are
\begin{equation}
 \hat{H}_{HA}=\sum_{i=1}^{2}\set{\frac{1}{2}\sbr{\hat{\mbf{p}}_{i}+\frac{1}{c}\mbf{A}\rbr{\mbf{r}_{i}}}^{2}+\frac{1}{2}\omega_{0}^{2}r_{i}^{2}}+\frac{1}{\abs{\mbf{r}_{2}-\mbf{r}_{1}}}, \label{Hooke_H}\\
\end{equation}
and
\begin{equation}
 \hat{H}_{ISI}=\sum_{i=1}^{2}\set{\frac{1}{2}\sbr{\hat{\mbf{p}}_{i}+\frac{1}{c}\mbf{A}\rbr{\mbf{r}_{i}}}^{2}+\frac{1}{2}\omega_{0}^{2}r_{i}^{2}}+\frac{\alpha}{\rbr{\mbf{r}_{1}-\mbf{r}_{2}}^{2}}. \label{ISI_H}
\end{equation}
Here, $\mbf{A}=\frac{1}{2}\rbr{\mbf{B}\times\mbf{r}}$ in the symmetric gauge, which is of the form of Eq.~(\ref{gauges}). Following Ref.~\cite{Vignale1987}, we
neglect spin terms in the Hamiltonians to concentrate on the features of the orbital currents.
For the ISI system, we can solve the time-independent Schr\"{o}dinger equation exactly for all frequencies and values of $m$ and $\alpha$. However, for Hooke's Atom,
analytical solutions only exist for a discrete set of frequencies. In order to give us freedom over the frequencies we choose, we solve the
Schr\"odinger equation with the method in Ref.~\cite{Coe2008}, which allows us to numerically determine the solution with high precision for all frequencies.
\begin{figure}[t]
 \includegraphics[width=\columnwidth]{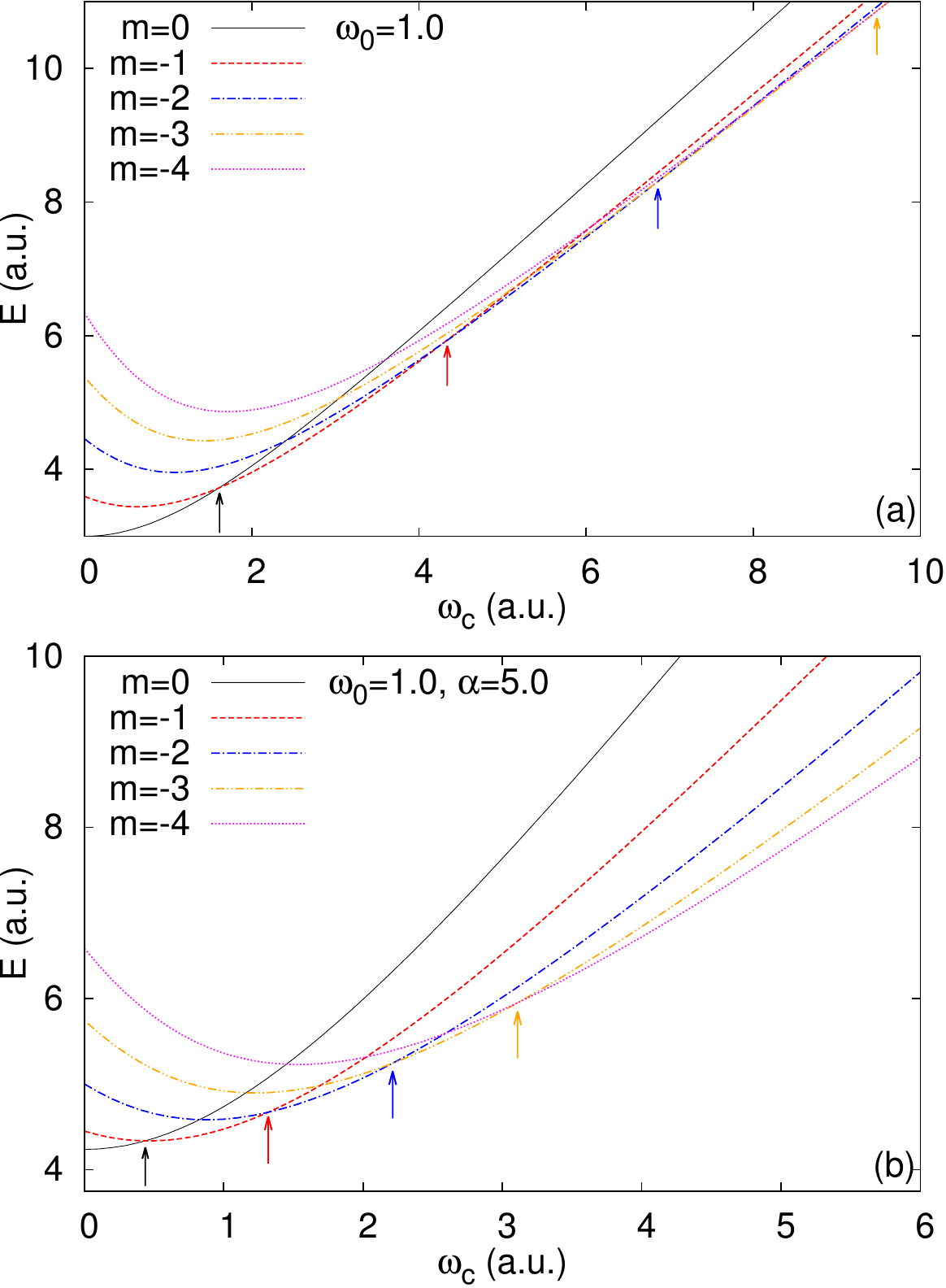}
 \caption{(Color online) Energy plotted against the cyclotron frequency for several values of $m$ for (a) Hooke's Atom and (b) the ISI system.
 The confinement frequency and interaction strength are held constant. Arrows indicate where the value of $m$ for the ground-state changes.}
 \label{energy}
\end{figure}

We generate families of ground states by varying the magnetic field via the cyclotron frequency $\omega_c$, while holding the confinement frequency, $\omega_0$,
and all other parameters in the Hamiltonian constant. For each $\omega_c$ value we calculate the wavefunction, particle density, and paramagnetic current
density. Within each family, one value of $\omega_{c}$ (and hence $m$) is selected as a reference ($\omega_{c_{ref}}$ and $m_{ref}$ respectively),
with the appropriate metrics used to find the distance between the physical functions at the reference and all of the other states in the family.
We choose the reference so that most of the available distance range is explored for both increasing and decreasing $\omega_c$.
\begin{figure*}[t]
 \includegraphics[width=\textwidth]{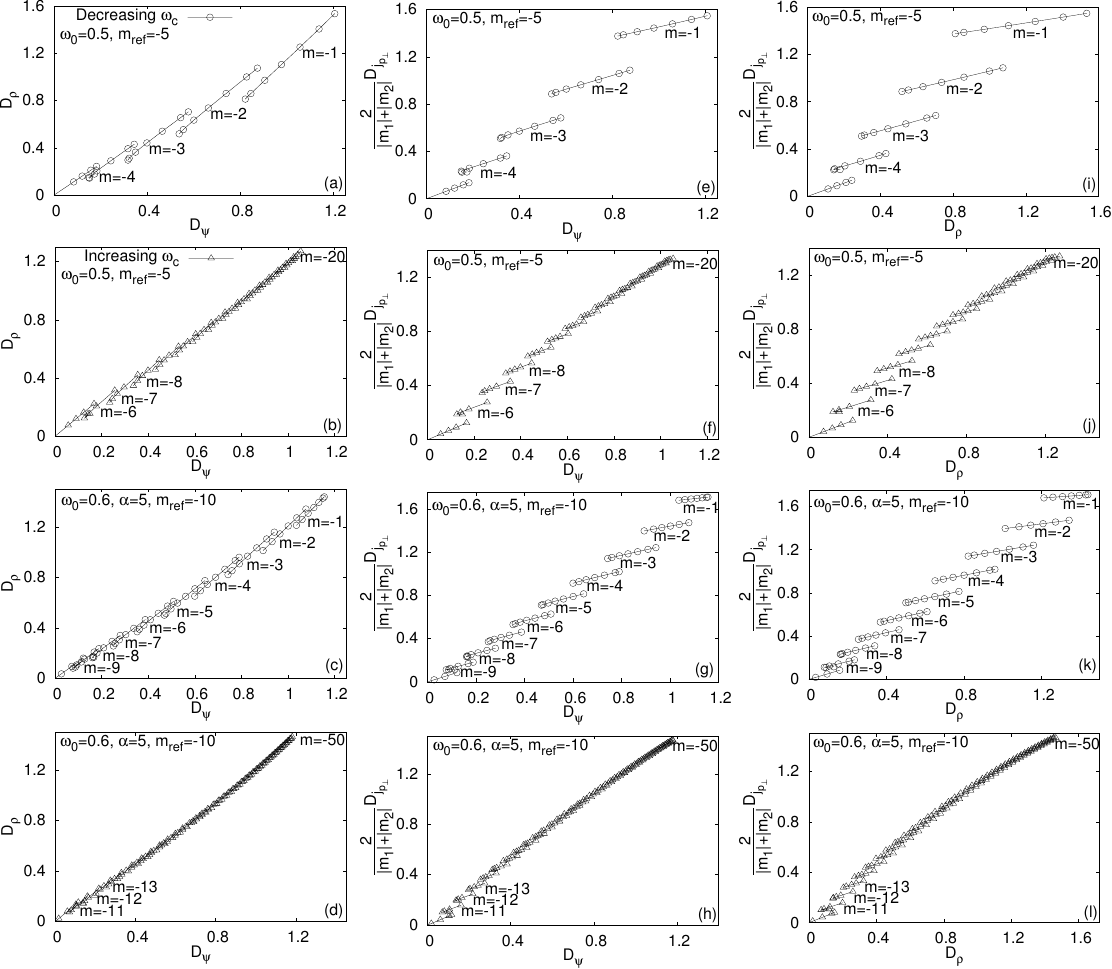}
 \caption{Plots of distances for Hooke's Atom with reference state ${\omega}_0=0.5, {\omega}_c=5.238,$ and $m_{ref}=-5$ (top two rows) and for the ISI system
 with reference state ${\omega}_0=0.6, {\omega}_c=5.36, {\alpha}=5,$ and $m_{ref}=-10$ (bottom two rows): (a) - (d) particle density distance
 against wavefunction distance, (e) - (h) paramagnetic current density distance against wavefunction distance, and (i) - (l) paramagnetic
 current density distance against particle density distance. The reference frequency is taken halfway between the two ``transition frequencies'' related to 
 $m_{ref}$.}
 \label{distances}
\end{figure*}

Figure~\ref{energy} shows
that, for both of our systems, the value of $m$ for which the energy is lowest decreases from zero through the negative integers as $\omega_{c}$ increases. Consequently, when
studying ground states, we must consider states on different spheres in the paramagnetic current density metric space.
We also note that there are ``transition frequencies'', i.e., values of $\omega_{c}$ where the energy is equal for two consecutive values of $m$.
These are the crossings of the energy curves in Fig.~\ref{energy}. Therefore, when varying $\omega_c$ it is necessary to change the value of $m$ at the
``transition frequencies'' in order to continue analysing ground states. Additionally, when $m=0$, $\mbf{j}_{p}(\mbf{r})=0$ for all $\mbf{r}$.
Hence, we take only negative values of $m$ to ensure we consider ground states with nonzero paramagnetic current densities.

\begin{figure}[t]
 \begin{center}
 \includegraphics[width=\columnwidth]{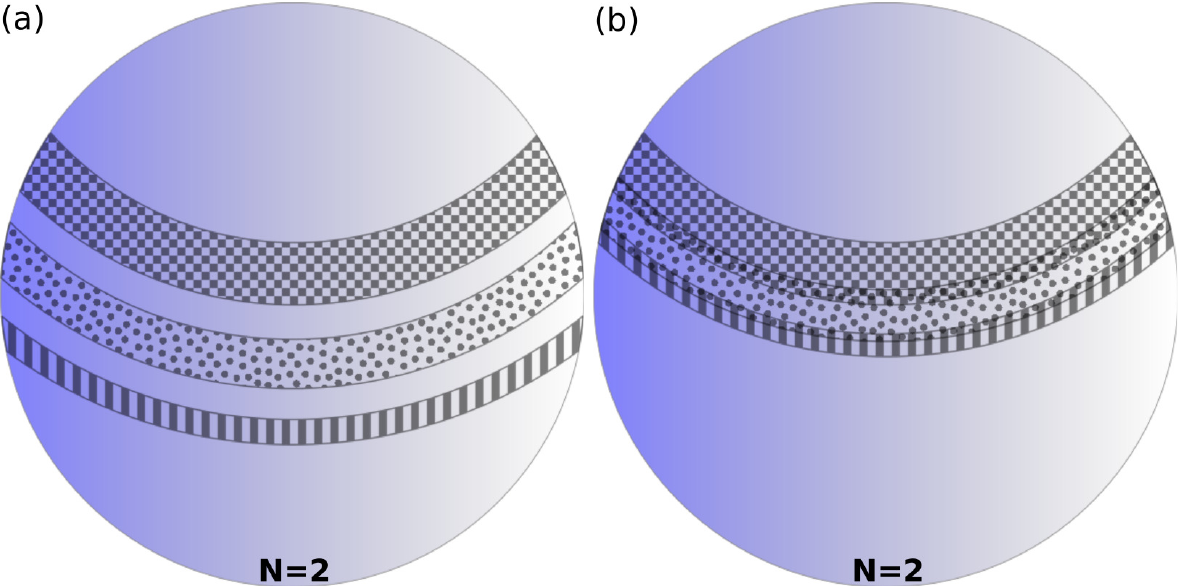}
 \caption{(Color online) Sketches of ``band structures'' consisting of (a) ``bands'' and ``gaps'' and (b) ``overlapping bands'' in particle density
 metric space for three consecutive bands, where a different patterning corresponds to a different value of $m$. The reference state is at the north pole.}
 \label{bands}
 \end{center}
\end{figure}

\subsection{Ground states' band structure and relevance to current density functional theory}

An important research area where properties of ground states are central is DFT. This theory has produced widely used tools
for realistic calculations of properties of many-body systems, such as band structures of metals and semiconductors, crystal structures of solids, and
characterisation of nanostructures~\cite{Capelle2006,Ullrich2013}. DFT is founded upon the Hohenberg-Kohn (HK) theorem \cite{HK1964}, which states
that there is a one-to-one mapping between ground-state wavefunctions and ground state particle densities.
There are various forms of DFT that
extend the application of the theory to a greater range of systems. CDFT is the extension of standard DFT to include systems subject to
external magnetic fields \cite{Vignale1987}. There is a HK-like theorem at the core of CDFT that states that a one-to-one mapping exists between the ground-state wavefunction
$\psi$ and, taken together, the particle density $\rho\rbr{\mbf{r}}$ and the paramagnetic current density $\mbf{j}_{p}\rbr{\mbf{r}}$
(CDFT-HK theorem) \cite{Vignale1987}. This additional complexity of the CDFT-HK theorem with respect to the original HK theorem is due to the fact that systems
with magnetic fields are characterised not only by a scalar potential (the external potential), but also by the vector potential connected
to the magnetic field \cite{Vignale1987}. In Ref.~\cite{Sharp2014} we started examining the CDFT-HK mapping by looking at the effect of varying the scalar potential, i.e., the
external confining potential; here we wish to complete this analysis by looking at the effect on the mapping of varying the vector potential, i.e., the magnetic field.

We start by comparing the distances between wavefunctions, their related particle densities, and
their related paramagnetic current densities. Figure~\ref{distances} shows plots of the relationships between the various distances considered, with
each point referring to a particular value of $\omega_c$.
Let us consider first the plots of particle density distance against wavefunction distance [Figs.~\ref{distances}(a) - \ref{distances}(d)]. As observed in Ref.~\cite{Sharp2014},
metric space regions corresponding to ground states present a ``band structure'', where points associated with the same value of $\abs{m}$ are grouped into distinct segments, i.e.,
bands. However, in contrast to the band structure observed in Ref.~\cite{Sharp2014} [sketched in Fig.~\ref{bands}(a)], when varying the vector potential we obtain a series of
``overlapping bands'', where the minimum wavefunction and minimum particle density distances for one value of $\abs{m}$ are smaller than the maximum distances for the previous value of
$\abs{m}$. This implies that there is an overlap between the projections of the bands on the metric space sphere representing the densities, as sketched in Fig.~\ref{bands}(b)
(similarly for the projection on the sphere representing the wavefunctions). Though overlapping, this band structure still results in discontinuities in the
relationship between $D_{\rho}$ and $D_{\psi}$ when the value of $m$ changes. Unlike when varying $\omega_0$ \cite{Sharp2014}, by varying the magnetic
field we do not observe any forbidden distances, so we cannot identify forbidden regions for ground states by considering the particle density and
wavefunction metric spaces alone. In the range of distances explored here, nearby wavefunctions are mapped onto nearby particle
densities and distant wavefunctions are mapped onto distant particle densities. However, in contrast to Ref.~\cite{Sharp2014}, the mapping is only piecewise linear:
When acting on the vector potential, as $\omega_c$ is swept through each transition frequency, ground states and their particle densities abruptly revert
to be closer to the reference state, while an almost linear mapping is maintained within two consecutive transition frequencies. The segments created in
this way do not overlap, as, at each transition frequency, the ball related to the particle density and centered at the reference density shrinks proportionally more than the corresponding ball
related to the wavefunction.
Also, in contrast with Ref.~\cite{Sharp2014}, the two families of ground states corresponding to $\abs{m}<\abs{m_{ref}}$ and $\abs{m}>\abs{m_{ref}}$ describe distinct
paths in metric space [e.g., compare Figs.~\ref{distances}(a) and \ref{distances}(b)], with the size of the bands greater for $\abs{m}<\abs{m_{ref}}$
compared to $\abs{m}>\abs{m_{ref}}$. For all of these reasons the CDFT-HK mapping between wavefunctions and related particle densities acquires added
complexity when varying the vector potential compared to varying the scalar potential [compare Figs.~2(a) and 2(b) in Ref.~\cite{Sharp2014} with Figs.~\ref{distances}(a) - \ref{distances}(d)].
\begin{figure*}[t]
 \includegraphics[width=\textwidth]{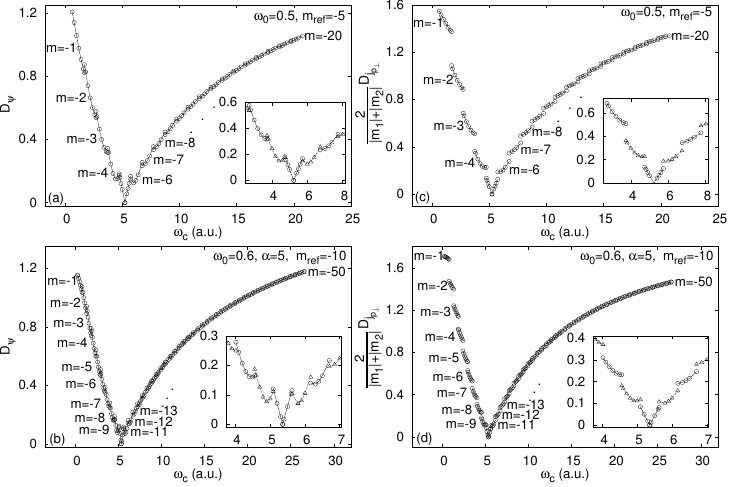}
 \caption{For Hooke's Atom (top) and the ISI system (bottom), (a) and (b) wavefunction distance and (c) and (d) paramagnetic current density distance
 are plotted against $\omega_c$. The behavior around the reference frequency is shown in each inset. The reference states are
 ${\omega}_0=0.5, {\omega}_{c_{ref}}=5.238,$ and $m_{ref}=-5$ for Hooke's Atom and ${\omega}_0=0.6, {\omega}_{c_{ref}}=5.36, {\alpha}=5,$ and $m_{ref}=-10$ for the ISI system.}
 \label{single_bands}
\end{figure*}

In Figs.~\ref{distances}(e) - \ref{distances}(h) we consider paramagnetic current density distance against wavefunction distance. Here we find once more
an overlapping band structure for wavefunction distances; however a band structure with regions of allowed (bands) and forbidden (gaps)
distances is observed for paramagnetic current density distances. In contrast with the one sketched in Fig.~\ref{bands}(a),
in this case each band resides on a different sphere according to the value of $\abs{m}$ (the radius of the sphere). Transition frequencies are points
of discontinuity for both paramagnetic current density and wavefunction distances. As for Figs.~\ref{distances}(a) - \ref{distances}(d), the curves
for increasing and decreasing $\omega_c$ (and hence $\abs{m}$) do not overlap, with larger bands for small values of $\abs{m}$. Finally Figs.~\ref{distances}(i) - \ref{distances}(l)
present the plots of paramagnetic current density distance against particle density distance. These exhibit behavior similar to that in Figs.~\ref{distances}(e) - \ref{distances}(h).

The overlapping band structures observed in Fig.~\ref{distances} demonstrate that mappings between some of the distances considered here are
multivalued. This multivalued mapping does \emph{not} represent a contradiction of the CDFT-HK theorem as it is entirely possible to have distinct functions
at the same distance away from a reference. In particular, in terms of the ``onion-shell'' geometry, all states situated at the same polar angle and on the
same sphere will have the same distance from the reference state.

\begin{figure}[t]
 \includegraphics[width=\columnwidth]{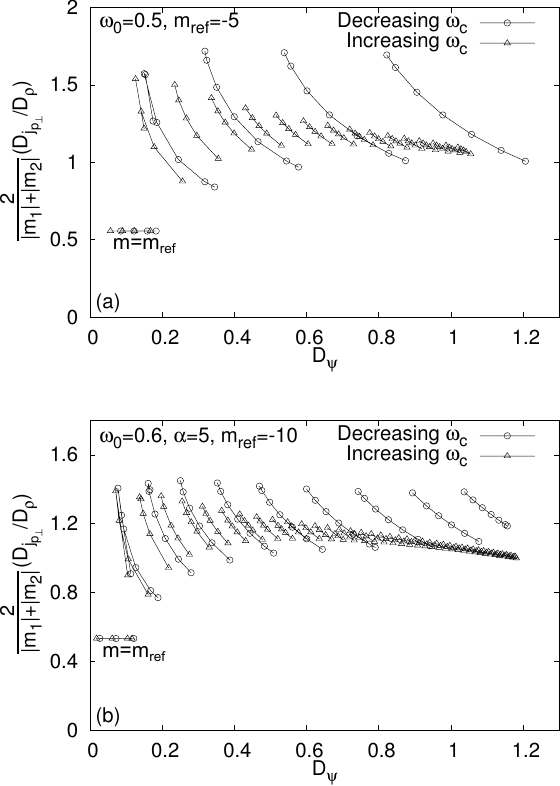}
 \caption{Plots of the ratio of paramagnetic current density distance to particle density distance against wavefunction distance for (a) Hooke's Atom with
 reference state ${\omega}_0=0.5, {\omega}_{c_{ref}}=5.238,$ and $m_{ref}=-5$, and (b) the ISI system with reference state
 ${\omega}_0=0.6, {\omega}_{c_{ref}}=5.36, {\alpha}=5,$ and $m_{ref}=-10$.}
 \label{gs_ratios}
\end{figure}

In Fig.~\ref{single_bands} the wavefunction and paramagnetic current density distances are plotted against $\omega_{c}$ for both systems, enabling
the band structures for individual functions to be analysed. We note that, as observed in Fig.~\ref{distances}, there is a decrease in the
wavefunction distance at transitions [Figs.~\ref{single_bands}(a) and \ref{single_bands}(b)], but an increase in the paramagnetic current density distance
[Figs.~\ref{single_bands}(c) and \ref{single_bands}(d)]. These features give rise to overlapping-band and band-gap structures, respectively. The other
major feature is that, when varying $\omega_c$, there is nonmonotonic behavior within bands corresponding to values of $m$ close to $m_{ref}$ (see insets).
For both wavefunctions and paramagnetic current densities, we observe that immediately after each transition frequency, the distances initially decrease to a
minimum for that particular band before increasing to the maximum for the band. This occurs at the
transition frequency to the next band. This behavior is more pronounced for wavefunctions than for paramagnetic current densities. As stated, the nonmonotonicity
is not in contradiction with the HK-like theorem of CDFT, but shows a richer behavior with respect to what was observed in Ref.~\cite{Sharp2014} when varying the scalar (confining) potential.

We point out that the band structure in metric space for paramagnetic current density is fundamentally different from the ones for particle density and wavefunction, as the
former develops on different spheres, one band for each sphere, while the latter are within a single sphere
where they may display overlapping-band or band-gap structures (see Fig.~\ref{bands}). All these band structures originate from the conservation law characterising
the paramagnetic current density and the features of the metric spaces for wavefunctions and particle densities are a direct consequence of the mapping of
$\mbf{j}_{p}\rbr{\mbf{r}}$ onto $\psi\rbr{\mbf{r}}$ and onto $\rho\rbr{\mbf{r}}$. In this sense the band structure features of the metric spaces for
wavefunctions and particle densities could be seen merely as the projections done by these mappings of the band structure characterising the paramagnetic current density.

\begin{figure*}[t]
 \includegraphics[width=\textwidth]{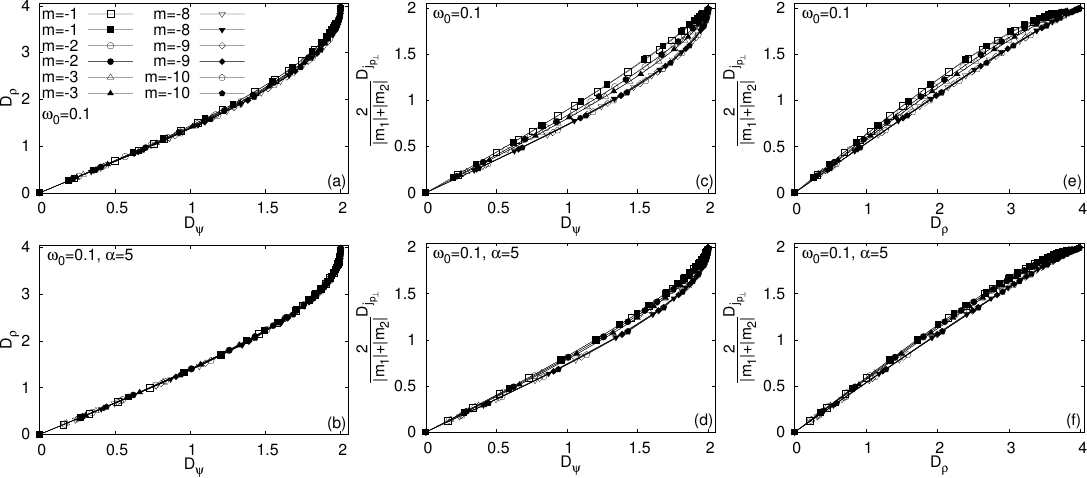}
 \caption{Plots of: (a) and (b) particle density distance against wavefunction distance, (c) and (d) paramagnetic current density distance against
 wavefunction distance, and (e) and (f) paramagnetic current density distance against particle density distance for $m=-1,-2,-3,-8,-9,-10$. The
 reference states, for each value of $m$, are: ${\omega}_0=0.1$ and ${\omega}_{c_{ref}}=30.0$ for Hooke's Atom (top) and ${\omega}_{0}=0.1,
 {\omega}_{c_{ref}}=5.0,$ and ${\alpha}=5$ for the ISI system (bottom). Closed symbols represent decreasing ${\omega}_c$ and open symbols represent increasing ${\omega}_c$.}
 \label{fixed_m}
\end{figure*}

Finally, we wish to concentrate on the implications of our findings for CDFT.
CDFT requires that both $\rho$ and $\mbf{j}_{p}$ are taken together to ensure a one-to-one mapping to the wavefunction.
The metric analysis allows us to provide evidence for an important aspect of this mapping, that is, to understand when the inclusion of
$\mbf{j}_{p}$ in the mapping becomes really crucial for the one-to-one correspondence to hold.

We present in Fig.~\ref{gs_ratios} the ratio $D_{\mbf{j}_{p_{\perp}}}/ D_{\rho}$ against $D_{\psi}$ for both Hooke's atom and the ISI system. From the
data it is immediately clear that, in metric space, to a good level of approximation, $D_{\mbf{j}_{p_{\perp}}}= const \times D_{\rho}$ as long as
$m=m_{ref}$. This constant is the same for
$\omega_{c}>\omega_{c_{ref}}$ and
$\omega_{c}<\omega_{c_{ref}}$. These findings suggest that, at least for
 the systems at hand, as long as we remain on the same sphere in the paramagnetic current density metric space,
$\mbf{j}_{p}$ and $\rho$ carry very similar information and the role of $\mbf{j}_{p}$ in the core mapping of CDFT is secondary.
The situation becomes very different for ground states with $m\neq m_{ref}$. In this case the ratio $D_{\mbf{j}_{p_{\perp}}}/D_{\rho}$ is far from
constant and Fig.~\ref{gs_ratios} clearly shows that the information contents of $\mbf{j}_{p}$ and $\rho$ are both necessary to define the state.
Similar results are obtained when keeping the magnetic field fixed but varying the confinement $\omega_0$ of the systems (not shown).

The characterisation of this difference in the role of $\mbf{j}_{p}$ and $\rho$ in the CDFT core mapping constitutes one of the main results of the paper.
To support it, we will analyse in the next section the behavior of states where $m$ is kept equal to $m_{ref}$ at all values of $\omega_c$.

\section{Excited States} \label{excite}

Although an understanding of the ground state is important for studying systems subject to magnetic fields, it is often necessary to go beyond
ground states, for example, when studying rapidly varying fields or spintronic devices that operate with excited states. With the metrics at hand, we
investigate excited states and consider distances between families of states corresponding to fixed values of $m$ \footnote{The center-of-mass
quantum number, $M$, is held constant at zero throughout this analysis.}. For each value of $m$ we will construct a family of states by varying $\omega_c$ ($\omega_0$ and $\alpha$ kept constant)
and calculating the corresponding wavefunctions, particle densities, and paramagnetic current densities. As for ground states, we choose $m<0$.
With respect to Fig.~\ref{energy}, this corresponds to following single energy curves smoothly, i.e., without switching to a different curve at crossings, as done instead for the ground state case.
Each family of states will then lie on a particular sphere in the paramagnetic current density metric
space. As the states considered are not necessarily ground states (see Fig.~\ref{energy}), there is no one-to-one mapping between the
wavefunction and particle and paramagnetic current densities, but,
these being fundamental quantities that characterise the system, we will still
explore their relationships. Additionally, the study of these quantities allows us to corroborate the findings related to Fig.~\ref{gs_ratios}.

\begin{figure}[b]
 \begin{center}
 \includegraphics[width=\columnwidth]{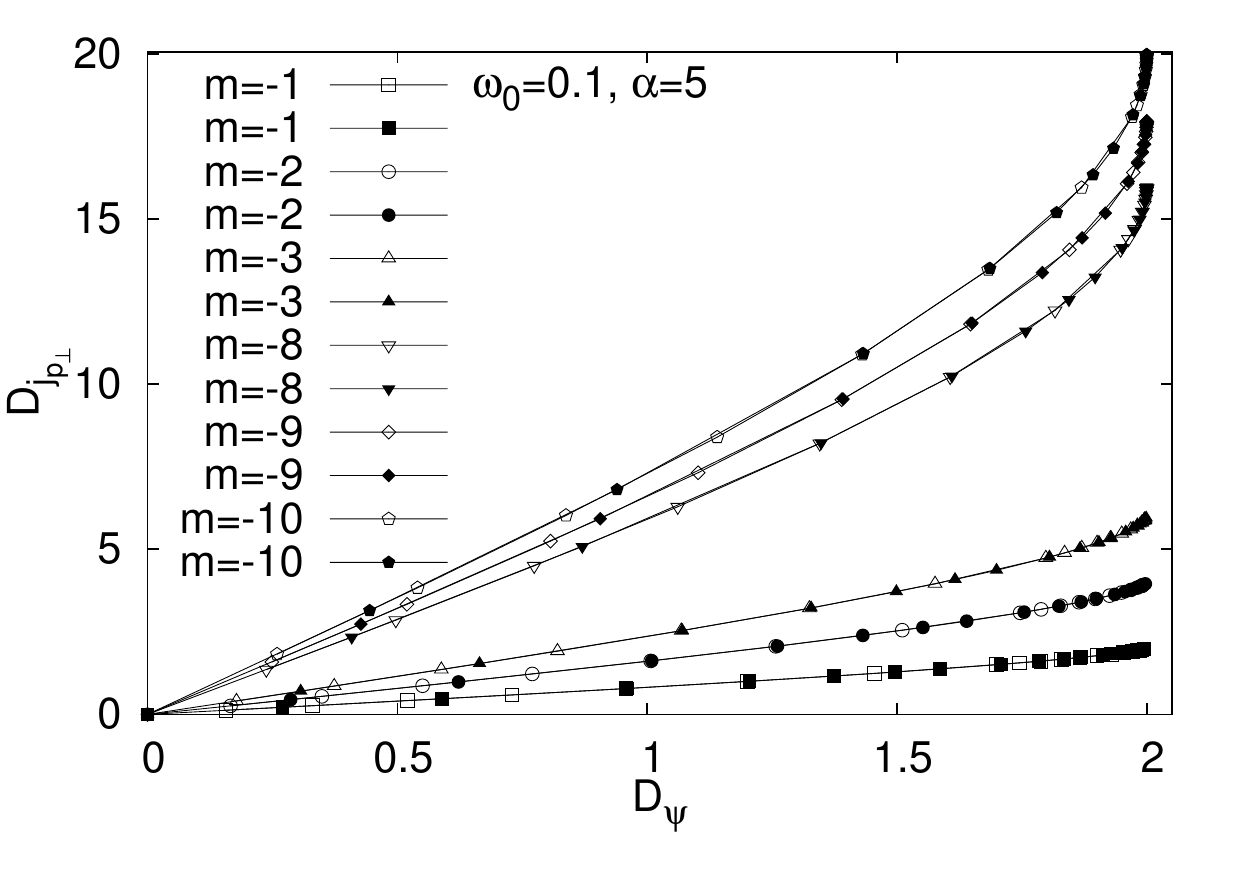}
 \caption{Plot of paramagnetic current density distance against wavefunction distance for $m=-1,-2,-3,-8,-9,-10$ for the ISI system. We take the
 state with $\omega_{0}=0.1, \omega_{c_{ref}}=5.0,$ and $\alpha=5$ as a reference for each value of $m$ and consider distances across the surface of each
 individual sphere.}
 \label{unscaled}
 \end{center}
\end{figure}

Figure~\ref{fixed_m} shows the relationship between each pair of distances for six different values of $m$.
For any pair of distances here discussed, we find a monotonic relationship that is linear in the short- to intermediate-distance regime, before one of the two functions rises more
sharply to its maximum (see also Fig.~\ref{unscaled}). The mapping between the physical functions is such that nearby functions $a$ (e.g., the wavefunctions) are mapped onto nearby functions $b$
(e.g., the paramagnetic current densities) and distant functions $a$ are mapped onto
distant functions $b$. Crucially, as opposed to ground states, distances do not form any kind of metric space ``band structure'', confirming the origin of band structures as the changes in $m$.

Looking at wavefunction distances against
particle density distances in Figs.~\ref{fixed_m}(a) and \ref{fixed_m}(b), and contrasting with Figs.~\ref{distances}(a)-\ref{distances}(d), we observe that
the curves for increasing and decreasing ${\omega}_c$ and all values of $\abs{m}$ collapse onto one another. This hints at a universal behavior
of the mapping between particle density and wavefunction when all the physical quantities describing the system remain on the same sphere in the related metric space while a physical parameter is smoothly changed.

\begin{figure}[b]
 \begin{center}
 \includegraphics[width=\columnwidth]{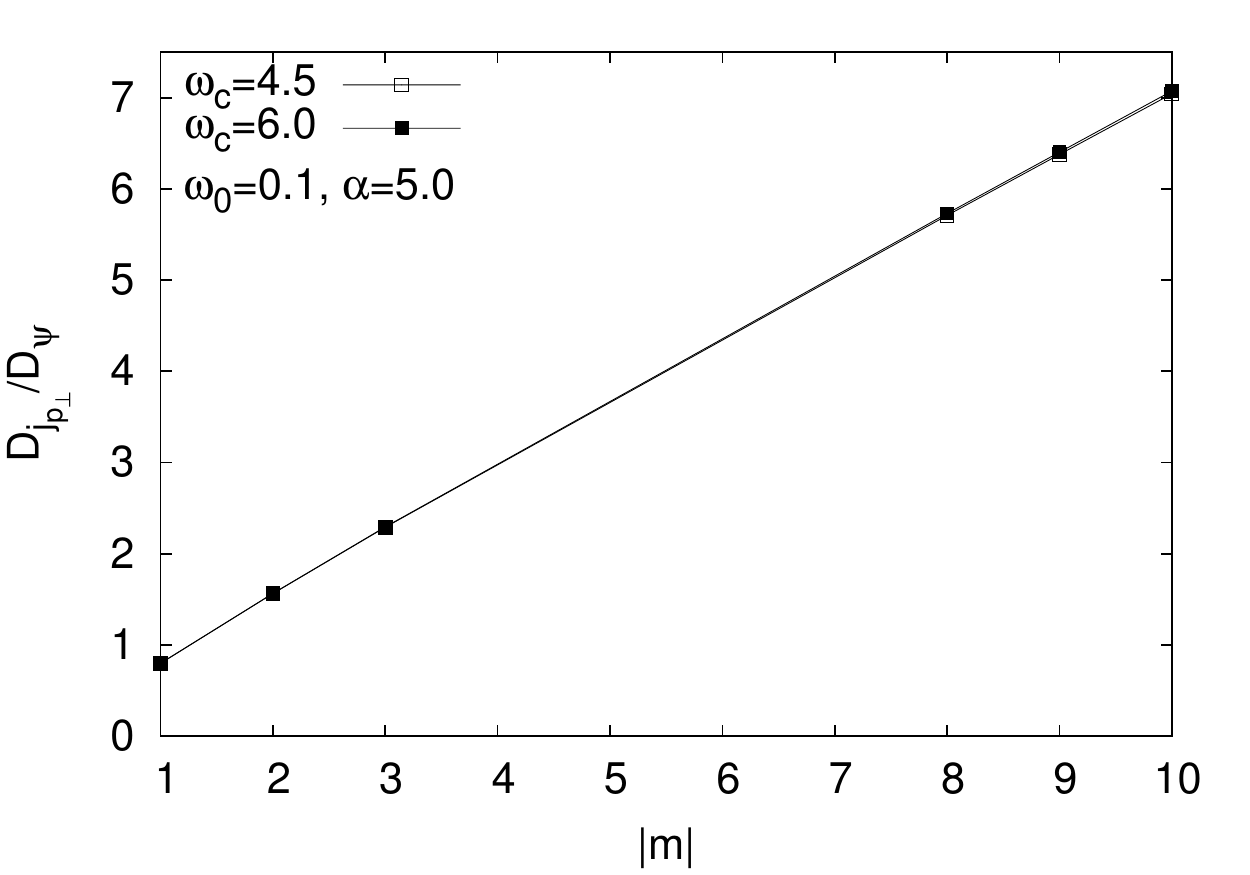}
 \caption{Plot of the ratio of paramagnetic current density distance to wavefunction distance against $\abs{m}$ for the ISI system. The reference
 for each value of $m$ is $\omega_{0}=0.1, \omega_{c_{ref}}=5.0, \alpha=5$, and the gradient is taken at $\omega_{c}=4.5$ for decreasing frequencies and
 $\omega_{c}=6.0$ for increasing frequencies, i.e., the frequencies corresponding to the closest points to $\omega_{c_{ref}}$ for
 both decreasing and increasing $\omega_{c}$ in Fig.~\ref{unscaled}.}
 \label{first_points}
 \end{center}
\end{figure}

When considering paramagnetic current density distance against wavefunction distance in Figs.~\ref{fixed_m}(c) and \ref{fixed_m}(d),
although the curves for increasing and decreasing $\omega_c$ collapse onto one another, the curves
for different values of $m$ are distinct, particularly when $\abs{m}$ is low. For lower values of
$\abs{m}$ the linear region extends across a larger range of distances. There is also a relatively
small increase in the gradient at greater distances for low $\abs{m}$. The curves in Figs.~\ref{fixed_m}(c) and \ref{fixed_m}(d) all
start and end at the same points. With the rescaling for
$D_{\mbf{j}_{p_{\perp}}}$ used in Fig.~\ref{fixed_m}, the curves tend to a limiting curve with increasing values of $\abs{m}$. In Fig.~\ref{unscaled}
we show the relationship between wavefunction distance and
paramagnetic current density distance for the ISI system without rescaling $D_{\mbf{j}_{p_{\perp}}}$.
Here, the curves for each value of $\abs{m}$ intersect only at the origin, and each has a unique maximum of $2\abs{m}$ for the paramagnetic current
density distance. We observe that the gradient of the initial linear
region increases with $\abs{m}$. Figure~\ref{first_points} shows, for
the ISI system, that the gradient
in this region increases linearly with $\abs{m}, D_{\mbf{j}_{p_{\perp}}} \approx k\abs{m}D_{\psi}$, with $0 \leqslant k\leqslant 1$,
and is approximately equal for both decreasing and increasing $\omega_{c}$. Similar results are obtained for Hooke's Atom (not shown).
These results imply that when rescaled as in Fig.~\ref{fixed_m}, the initial slope of the curves will always be below $45^{\circ}$, a result also
observed in Ref.~\cite{D'Amico2011} for the case in which different spheres in the wavefunction metric space geometry were considered.

When considering paramagnetic current density distance against particle density distance [Figs.~\ref{fixed_m}(e) and \ref{fixed_m}(f)] we see that, as for $D_{\mbf{j}_{p_{\perp}}}$ vs $D_{\psi}$,
with the rescaling of Fig.~\ref{fixed_m} there are distinct curves for each value of $m$ that converge onto a single curve as $\abs{m}$ increases.
As opposed to $D_{\mbf{j}_{p_{\perp}}}$ vs $D_{\psi}$, the extent of the linear behavior of these curves is increasing as $\abs{m}$ increases.

The behavior of the curves observed in Fig.~\ref{fixed_m} reflects the ``onion-shell'' geometry. For wavefunctions and particle densities
the sphere radius is associated with the number of particles in the
system, which is fixed for the systems considered. Thus, regardless of the value of $\abs{m}$, wavefunctions
and particle densities always lie on the same sphere in their
respective metric spaces. The fact that the related curves still superimpose for changing
$\abs{m}$ seems to imply that the value of $\abs{m}$ has no relevant effect on the curves representing the relative change of $\psi$ and $\rho$
for changing parameters, at least as long as they remain on the same sphere. In paramagnetic current density metric space, the spheres' radii are
related to $\abs{m}$, so paramagnetic current densities are on the surface of different spheres each time we consider a different value of $\abs{m}$.
As a result we see that the curves' shape is affected and they do not collapse onto each other. A similar universal behavior within each sphere
and, by contrast, the breaking of this universality when different spheres were considered, was also observed in Ref.~\cite{D'Amico2011}, where different
values of $N$, and hence different spheres, for both wavefunctions and particle densities were considered. This seems to suggest that different
behavior for the mappings should be expected when curves on different spheres in the metric spaces are involved.
\begin{figure}[t]
 \includegraphics[width=\columnwidth]{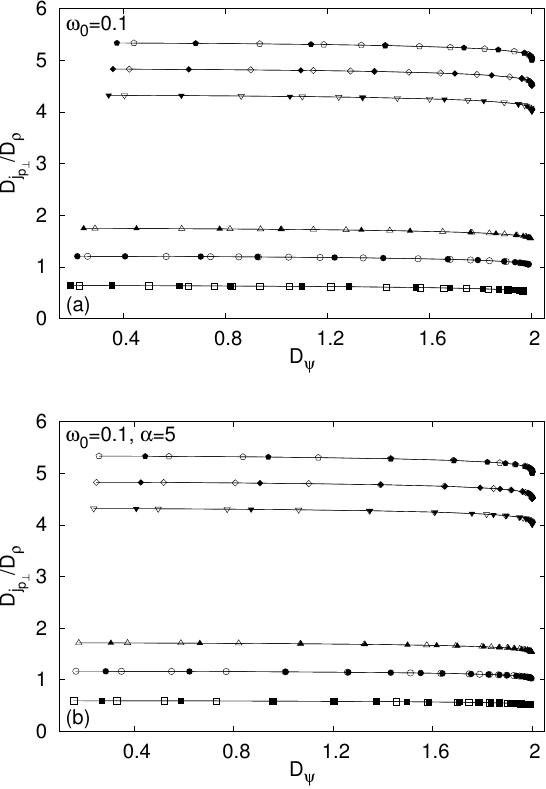}
 \caption{Plots of the ratio of paramagnetic current density distance to particle density distance against wavefunction distance for (a) Hooke's Atom, with reference state
 ${\omega}_0=0.1, {\omega}_{c_{ref}}=30.0$, and (b) the ISI system, with reference state ${\omega}_0=0.1, {\omega}_{c_{ref}}=5.0,$ and ${\alpha}=5$.
 Closed symbols represent decreasing ${\omega}_c$ and open symbols represent increasing ${\omega}_c$.}
 \label{fixed_m_ratio}
\end{figure}

Finally, Fig.~\ref{fixed_m_ratio} combines all distances for each system in a single plot. Importantly, this figure shows that for small to medium wavefunction distances
$D_{\mbf{j}_{p_{\perp}}}/D_{\rho}\sim constant$, where the constant depends on $\abs{m}$, so this ratio is, to a good approximation, independent over variations of the wavefunction
for relatively close wavefunctions. In this respect, for relatively close wavefunctions this suggests that the mappings between current density and wavefunction
and between particle density and wavefunction are very similar, as long as the family of states follows the evolution of the same energy
eigenstate as driven by the varying parameter (see Fig.~\ref{energy}).

\section{Conclusion} \label{conclusion}

The metric space approach to quantum mechanics has
enabled us to illustrate the role of the vector potential in systems
subject to external magnetic fields, with particular reference to the
fundamental concepts of CDFT. Importantly, we have also furthered the
theoretical framework of the metric space approach to quantum
mechanics by discussing the key point of gauge invariance for the
natural metrics proposed and in particular demonstrating the gauge
invariant metric for the paramagnetic current density, which is not a gauge invariant quantity \textit{per se}.

The presence of the vector potential in the Hamiltonian leads to the inclusion of the paramagnetic current density in the core mapping of CDFT. By
considering the metric for the paramagnetic current density together with that for the particle density, we were able to investigate the relative
contribution to this core mapping from each of these two quantities, for the systems at hand. When $m$ is held constant, and paramagnetic current
and particle densities belong to the same metric space sphere as their reference state, we observed that the ratio
$D_{\mbf{j}_{p_{\perp}}}/D_{\rho}$ is approximately constant, suggesting that $\rho$ and $\mbf{j}_{p}$ contribute similar information. However, this
simple relation dramatically breaks down when considering states with $m\ne m_{ref}$ and hence states spanning different spheres in paramagnetic current density metric space. This
suggests that the presence of $\mbf{j}_{p}$ in the core mapping of CDFT becomes crucial in this case.

By varying the vector potential, we uncovered different aspects of the ``band structure'' in ground-state
metric spaces, in particular the presence of overlapping bands, which
enriches the band-gap structures already observed when varying
the scalar potential. Our analysis suggests that, in general, the
presence of bands in metric space can be expected when considering a family of states for which one of the fundamental physical functions spans more than one sphere in its metric space.
For ground states, the onset of the band structure is the signature of energy levels' crossings obtained by varying a parameter in the Hamiltonian (the magnetic field in the present case).

We also applied the metric space approach to quantum mechanics beyond ground states. When considering families of states characterised by fixed values of $m$, it was found that the mappings
between wavefunctions, particle densities, and paramagnetic current densities are monotonic and almost linear, without band structures, confirming that each band is characterised by a specific value of $m$. The curves $D_{\psi}$ versus $D_{\rho}$ superimpose for all values of $m$, but not so when $D_{\mbf{j}_{p_{\perp}}}$ is involved. This is consistent with
the fact that a different $m$ represents different spheres in the ``onion-shell'' geometry related to $\mbf{j}_{p}$.

Finally, when considering the ratio $D_{\mbf{j}_{p_{\perp}}}/D_{\rho}$ for these ``fixed-$m$'' families, the relationship $D_{\mbf{j}_{p_{\perp}}}/D_{\rho}\approx const$ was observed to persist up to intermediate distances, and for all of the values of $m$ that
were explored. At least for the systems at hand, this suggests that, within the same sphere and up to quite different states, particle density and wavefunction still suffice to contribute most of the information on the physical system.

\begin{acknowledgments}
The authors gratefully acknowledge support from a University of York - FAPESP combined grant. P.~M.~S. acknowledges EPSRC for financial support.
I.~D. acknowledges support by CNPq Grant: PVE--Processo: 401414/2014-0. All data created during this research are available by request from the University of York Data Catalogue
http://dx.doi.org/10.15124/9a122b40-8ef0-435a-8c9e-4061c84c292f
\end{acknowledgments}

% Specify following sections are appendices. Use \appendix* if there is only one appendix.
\appendix

\section{Gauge invariance for the wavefunction metric} \label{wave_metric}

Gauge transformations affect wavefunctions by introducing a constant global phase factor [see Eq.~(\ref{psi_trans})].
Wavefunctions differing by this phase factor describe the same physics; in fact, the solutions of the Schr\"{o}dinger equation are only
defined up to a global phase factor. To have physically meaningful metrics, it is therefore important to define equivalence classes such
that the metric assigns zero distance to wavefunctions differing only by a {\it global} phase factor.

An equivalence class for an element $x\in X$ is defined as \cite{Kubrusly2009}
\begin{equation} \label{class}
 \sbr{x}=\set{x'\in X:x\sim x'},
\end{equation}
where $\sim$ is the equivalence relation. Each element of the set $X$ belongs to a single equivalence class \cite{Kubrusly2009}.

In order to account for an equivalence relation between elements, $x\sim x'$, we follow a general procedure for introducing equivalence relations into
a metric space $\rbr{X,D}$. We define the function \cite{Burago2001}
\begin{equation} \label{quot_semi}
 D_{R}(x,y)=\inf\set{\sum_{i=1}^{k} D(p_{i},q_{i}):p_{1}=x,q_{k}=y,k\in\mbb{N}},
\end{equation}
where the infimum is taken over all choices of $\set{p_{i}},$ and $\set{q_{i}}$ such that $p_{i+1}\sim q_{i}$. This implies that if $x\sim y$,
$D_{R}\rbr{x,y}=D\rbr{x,x}+D\rbr{y,y}=0$ even if $D\rbr{x,y}\neq0$ \cite{Burago2001}. This function is a semimetric (or pseudometric)
on the set $X$, known as the quotient semimetric. A semimetric is a
distance function that obeys all of the axioms of a metric except that
it allows zero distance between nonidentical elements as well as identical ones.

For wavefunctions, the metric derived from the conservation law before accounting for the equivalence of wavefunctions differing by a global phase is \cite{Longpre2008,D'Amico2011}
\begin{equation} \label{old_dpsi}
 \wt{D}_{\psi}\rbr{\psi_{1},\psi_{2}}=\sbr{\int\abs{\psi_{1}-\psi_{2}}^{2} d\mbf{r}_{1}\ldots d\mbf{r}_{N}}^{\frac{1}{2}}.
\end{equation}
For this in general we have that $ \wt{D}_{\psi}\rbr{\psi,e^{i\phi}\psi}\ne 0$.
If in Eq.~(\ref{quot_semi}) we take $k=2$, we find
\begin{align}
 D_{\psi}\rbr{\psi_{1},\psi_{2}}&=\inf\set{\wt{D}_{\psi}\rbr{\psi_{1},\psi'}+\wt{D}_{\psi}\rbr{\psi_{2},\psi_{2}}}\\
 &=\inf\set{\wt{D}_{\psi}\rbr{\psi_{1},\psi'}}.
\end{align}
where $\psi'=e^{i\phi}\psi_2\sim\psi_{2}$ and we have used the positivity axiom of the metric. The choice of $\psi'$ that will minimise the value of the semimetric is determined by the phase factor, hence
\begin{align} \label{dpsi_def}
 D_{\psi}\rbr{\psi_{1},\psi_{2}}&=\min_{\phi}\set{\wt{D}_{\psi}\rbr{\psi_{1},e^{i\phi}\psi_2}}
\end{align}
With this semimetric space $\rbr{\set{\psi},D_{\psi}}$, we can recover a metric space in a natural way, by ``gluing'' equivalent
elements to form a set of equivalence classes. By considering the set of equivalence classes, rather than the set of all wavefunctions,
all wavefunctions differing only by a global phase factor are identified with one another. Thus, for wavefunctions, the set of equivalent
wavefunctions with $D_{\psi}$ is a metric space, with the metric defined between each of the equivalence classes, as required
\cite{Kubrusly2009}. The metric $D_{\psi}$ defined from Eq.~(\ref{dpsi_def}) is the same as the metric defined in Refs.~\cite{Longpre2008,D'Amico2011}
and can be expressed in the form of Eq.~(\ref{dpsi}).

\section{Determining the gauges where $L_{z}$ is a constant of motion} \label{vector_potentials}

In order to be a constant of motion, the $z$ component of the angular momentum $\hat{L}_z=-i\sbr{\mbf{r}\times{\nabla}}_{z}$ must commute with the Hamiltonian.
Given that a vector potential is present, we consider the Pauli Hamiltonian
\begin{equation} \label{pauli_H}
 \hat{H}=-\frac{1}{2}\nabla^{2}+\frac{i\hbar e}{2m_{e}}\rbr{\mbf{A}\cdot\nabla+\nabla\cdot\mbf{A}+\frac{e^{2}}{2m_{e}}\mbf{A}^{2}}+V\rbr{\mbf{r}},
\end{equation}
with $V\rbr{\mbf{r}}$ such that $\sbr{V\rbr{\mbf{r}},\hat{L}_{z}}=0$. The Hamiltonian~(\ref{pauli_H}) does not necessarily commute with $\hat{L}_z$ for a particular
$\mbf{A}\rbr{\mbf{r}}$, because $\hat{L}_{z}$ is gauge variant. For instance, $\hat{L}_z$ commutes with the Hamiltonian~(\ref{pauli_H}) in the symmetric gauge
$\mbf{A}=\sbr{y,-x,0}$ and does not commute with it in the Landau gauge $\mbf{A}=\sbr{0,-x,0}$. We wish to determine the general set of vector potentials
where $\sbr{\hat{H},\hat{L}_{z}}=0$. \footnote{In the case where we have many-body interactions, we only consider the case $\sbr{U\rbr{\mbf{r}_{i},\mbf{r}_{j}},\hat{L}_z}=0$,
as is the case for the Coulomb interaction.}

\subsection{Simplifying the commutator}

The commutator we wish to evaluate is
\begin{align} \label{commutator}
 \sbr{\hat{H},\hat{L}_z}\psi=&\mbf{A}\cdot\nabla\rbr{x\pdiff{\psi}{y}}-\mbf{A}\cdot\nabla\rbr{y\pdiff{\psi}{x}}-x\pdiff{}{y}\rbr{\mbf{A}\cdot\nabla\psi}\nonumber\\
 &+y\pdiff{}{x}\rbr{\mbf{A}\cdot\nabla\psi}+\nabla\cdot\rbr{\mbf{A}x\pdiff{\psi}{y}}-\nabla\cdot\rbr{\mbf{A}y\pdiff{\psi}{x}}\nonumber\\
 &-x\pdiff{}{y}\nabla\cdot\rbr{\mbf{A}\psi}+y\pdiff{}{x}\nabla\cdot\rbr{\mbf{A}\psi}+A^{2}x\pdiff{\psi}{y}\nonumber\\
 &-A^{2}y\pdiff{\psi}{x}-x\pdiff{}{y}\rbr{A^{2}\psi}+y\pdiff{}{x}\rbr{A^{2}\psi},
\end{align}
where we have used that $\sbr{\frac{-\hbar^{2}}{2m_{e}}\nabla^{2}+V\rbr{\mbf{r}},\hat{L}_{z}}=0$.
We wish to impose the condition $\sbr{\hat{H},\hat{L}_z}=0$, and then solve the commutator to obtain the
vector potential $\mbf{A}\rbr{\mbf{r}}$. After performing the vector operations and simplifying, Eq.~(\ref{commutator}) reduces to
\begin{align} \label{commute_eqn}
 &\rbr{2A_{x}\pdiff{\psi}{y}-2A_{y}\pdiff{\psi}{x}}-2x\rbr{\pdiff{\psi}{x}\pdiff{A_{x}}{y}+\pdiff{\psi}{y}\pdiff{A_{y}}{y}+\pdiff{\psi}{z}\pdiff{A_{z}}{y}}\nonumber\\
 &+2y\rbr{\pdiff{\psi}{x}\pdiff{A_{x}}{x}+\pdiff{\psi}{y}\pdiff{A_{y}}{x}+\pdiff{\psi}{z}\pdiff{A_{z}}{x}}\nonumber\\
 &-x\psi\pdiff{}{y}\rbr{\pdiff{A_{x}}{x}+\pdiff{A_{y}}{y}+\pdiff{A_{z}}{z}}\nonumber\\
 &+y\psi\pdiff{}{x}\rbr{\pdiff{A_{x}}{x}+\pdiff{A_{y}}{y}+\pdiff{A_{z}}{z}}\nonumber\\
 &-x\psi\pdiff{A^{2}}{y}+y\psi\pdiff{A^{2}}{x}=0.
\end{align}
In order to progress with the solution of this equation, we first consider the case where $\psi,\pdiff{\psi}{x},\pdiff{\psi}{y},$ and $\pdiff{\psi}{z}$ are all
independent of each other. This choice allows us to decompose Eq.~(\ref{commute_eqn}) into a set of simultaneous equations, which we can then solve.
The solution of these equations will provide properties of the general set of vector potentials where $\sbr{\hat{H},\hat{L}_{z}}=0$. Using these properties, we will then 
solve Eq.~(\ref{commute_eqn}) for $\mbf{A}\rbr{\mbf{r}}$ using a general wavefunction.

With our choice of trial wavefunction, we write the set of simultaneous equations
\begin{align}
 -&x\pdiff{}{y}\rbr{\pdiff{A_{x}}{x}+\pdiff{A_{y}}{y}+\pdiff{A_{z}}{z}}-x\pdiff{A^{2}}{y}\nonumber\\
 &+y\pdiff{}{x}\rbr{\pdiff{A_{x}}{x}+\pdiff{A_{y}}{y}+\pdiff{A_{z}}{z}}+y\pdiff{A^{2}}{x}=0,\label{sim_1}\\
 A&_{y}+x\pdiff{A_{x}}{y}-y\pdiff{A_{x}}{x}=0,\label{sim_2}\\
 A&_{x}-x\pdiff{A_{y}}{y}+y\pdiff{A_{y}}{x}=0,\label{sim_3}\\
 y&\pdiff{A_{z}}{x}-x\pdiff{A_{z}}{y}=0.\label{sim_4}
\end{align}
We concentrate first on Eqs.~(\ref{sim_2})-(\ref{sim_4}), a set of three equations for the three unknowns $A_{x}$, $A_{y}$, and $A_{z}$.
Firstly, we consider Eq.~(\ref{sim_4}): In order to solve this partial differential equation (PDE), we use the method of characteristics \cite{rhee2001}.

The method of characteristics requires the visualisation of
Eq.~(\ref{sim_4}) in four-dimensional coordinates $\rbr{x,y,z,u}$. By
considering the solution surface $u=A_{z}\rbr{x,y,z}$, we can write
\begin{equation*}
 A_{z}\rbr{x,y,z}-u=0.
\end{equation*}
For any surface, $S$, a normal vector to the surface is given by $\nabla S$. Thus, the vector
$\sbr{\pdiff{A_{z}}{x},\pdiff{A_{z}}{y},\pdiff{A_{z}}{z},-1}$ is normal to the solution surface.
We now write the PDE~(\ref{sim_4}) as a scalar product
\begin{equation*}
 \sbr{y,-x,0,0}\cdot\sbr{\pdiff{A_{z}}{x},\pdiff{A_{z}}{y},\pdiff{A_{z}}{z},-1}=0.
\end{equation*}
Since the scalar product of these two vectors is zero, they must be orthogonal. Given also that the vector
$\sbr{\pdiff{A_{z}}{x},\pdiff{A_{z}}{y},\pdiff{A_{z}}{z},-1}$ is normal to the surface, this tells us that
the vector field $\sbr{y,-x,0,0}$ is tangent to the surface at every point, providing a geometrical interpretation
of the PDE. Thus, any curve within the surface $A_{z}\rbr{x,y,z}-u=0$ that has the vector $\sbr{y,-x,0,0}$ as a tangent
at every point must lie entirely within the surface. Such curves are called characteristic curves \cite{rhee2001}.
Any curve can be described by a parameter $t$ and the tangent of such a curve $\mbf{r}\rbr{t}$ is given by the derivative with
respect to this parameter $\mbf{r}'\rbr{t}$. Therefore, the tangent of a characteristic curve
$\mbf{r}\rbr{t}=\sbr{x\rbr{t},y\rbr{t},z\rbr{t},A_{z}\rbr{t}}$ is given by the vector
\begin{equation*}
 \mbf{r}'\rbr{t}=\sbr{\diff{x}{t},\diff{y}{t},\diff{z}{t},\diff{A_{z}}{t}}
\end{equation*}
This vector is therefore proportional to the tangent vector $\sbr{y,-x,0,0}$ for this characteristic curve,
allowing us to construct the equations
\begin{align}
 \diff{x}{t}&=y,\label{c_ode_1}\\
 \diff{y}{t}&=-x,\label{c_ode_2}\\
 \diff{z}{t}&=0,\label{c_ode_4}\\
 \diff{A_{z}}{t}&=0.\label{c_ode_3}
\end{align}
These are the characteristic equations of the PDE~(\ref{sim_4}).

Solving this set of ordinary differential equations (ODEs) yields the solution of the original PDE~(\ref{sim_4}), since
\begin{align*}
 \diff{A_{z}}{t}&=\diff{x}{t}\pdiff{A_{z}}{x}+\diff{y}{t}\pdiff{A_{z}}{y}+\diff{z}{t}\pdiff{A_{z}}{z}\\
 &=y\pdiff{A_{z}}{x}-x\pdiff{A_{z}}{y}=0.
\end{align*}
By eliminating the parameter $t$ in Eqs.~(\ref{c_ode_1})-(\ref{c_ode_3}), we can reduce the set of ODEs to three equations
\begin{align}
 \diff{y}{x}&=-\frac{x}{y},\label{ode_1}\\
 \diff{z}{x}&=0,\label{ode_2}\\
 \diff{A_{z}}{x}&=0.\label{ode_3}
\end{align}
We now note that the constant of integration in Eq.~(\ref{ode_3}) has a functional dependence on the solutions to Eqs.~(\ref{ode_1}) and (\ref{ode_2}).
This is because the ODEs are solved along characteristic curves: The constants of integration are constant along a particular characteristic, but
can vary between characteristics. The solutions to Eqs.~(\ref{ode_1}) and (\ref{ode_2}) are
\begin{equation}\label{char_solns}
 x^{2}+y^{2}=a \qquad z=b,
\end{equation}
respectively, where $a$ and $b$ are the constants of integration. Thus, the solution for $A_{z}$ is
\begin{equation}\label{A_z_soln}
 A_{z}=\gamma\rbr{{x^{2}+y^{2},z}},
\end{equation}
where $\gamma$ is an arbitrary function.

\subsection{Solving the simultaneous equations}

We will now solve Eqs.~(\ref{sim_2}) and (\ref{sim_3}) simultaneously. First, we differentiate Eq.~(\ref{sim_2}) with respect to both $x$ and $y$, which gives
\begin{align}
 \pdiff{A_{y}}{x}+\pdiff{A_{x}}{y}+x\pmixdiff{A_{x}}{x}{y}-y\pdoublediff{A_{x}}{x}=0,\\
 \pdiff{A_{y}}{y}+x\pdoublediff{A_{x}}{y}-\pdiff{A_{x}}{x}-y\pmixdiff{A_{x}}{x}{y}=0,
\end{align}
respectively. We substitute these expressions for $\pdiff{A_{y}}{x}$ and $\pdiff{A_{y}}{y}$ into Eq.~(\ref{sim_3}) and obtain
\begin{align} \label{big_pde}
 A_{x}&-x\rbr{-x\pdoublediff{A_{x}}{y}+\pdiff{A_{x}}{x}+y\pmixdiff{A_{x}}{x}{y}}\nonumber\\
 &+y\rbr{-\pdiff{A_{x}}{y}-x\pmixdiff{A_{x}}{x}{y}+y\pdoublediff{A_{x}}{x}}=0,\nonumber\\
 y&^{2}\pdoublediff{A_{x}}{x}-2xy\pmixdiff{A_{x}}{x}{y}+x^{2}\pdoublediff{A_{x}}{y}-x\pdiff{A_{x}}{x}-y\pdiff{A_{x}}{y}+A_{x}=0.
\end{align}
We now have an equation containing only the unknown $A_{x}$ that we can solve.

We begin to solve this equation by using the method of characteristics. For second order PDEs, it is first necessary to determine the
type of the PDE, either hyperbolic, parabolic, or elliptic. This is done by calculating the discriminant $b^{2}-4ac$, where
$a,b,$ and $c$ are the coefficients of $\pdoublediff{A_{x}}{x},\pmixdiff{A_{x}}{x}{y},$ and $\pdoublediff{A_{x}}{y}$ respectively. This will then allow us
to perform an appropriate change of variables from $\rbr{x,y,z}$ to $\rbr{\xi,\eta,z}$, where $\xi$ and $\eta$ are the characteristics \cite{Kreyzig2006}.
The discriminant is
\begin{equation}
 b^{2}-4ac=4x^{2}y^{2}-4x^{2}y^{2}=0.
\end{equation}
Therefore, the characteristic equation is parabolic and has one repeated solution, which we take for $\xi$.
The characteristic equation is the ODE \cite{Kreyzig2006}
\begin{equation} \label{char}
 y^{2}\rbr{\diff{y}{x}}^{2}-2xy\diff{y}{x}+x^{2}=0;
\end{equation}
solving for $\diff{y}{x}$ gives
\begin{equation*}
 \diff{y}{x}=-\frac{x}{y}.
\end{equation*}
Hence, from Eq.~(\ref{char_solns}) we know that the first characteristic is $\xi=a=x^{2}+y^{2}$. Since there is only one root of the characteristic equation,
we have complete freedom in the choice for $\eta$, provided that it is not the same as $\xi$. Given that we know that the symmetric gauge satisfies the
commutator $\sbr{\hat{H},\hat{L}_z}=0$ and, specifically, that $A_{x}=y$ is a solution for Eq.~(\ref{big_pde}), we choose $\eta=y$.
By using the chain rule, we find the derivatives in Eq.~(\ref{big_pde}),
\begin{align*}
 \pdiff{A_{x}}{x}&=2x\pdiff{A_{x}}{\xi},\\%\label{derivative_x}\\
 \pdiff{A_{x}}{y}&=2y\pdiff{A_{x}}{\xi}+\pdiff{A_{x}}{\eta},\\%\label{derivative_y}\\
 \pdoublediff{A_{x}}{x}&=4x^{2}\pdoublediff{A_{x}}{\xi}+2\pdiff{A_{x}}{\xi},\\%\label{derivative_xx}\\
 \pdoublediff{A_{x}}{y}&=4y^{2}\pdoublediff{A_{x}}{\xi}+4y\pmixdiff{A_{x}}{\xi}{\eta}+\pdoublediff{A_{x}}{\eta}+2\pdiff{A_{x}}{\xi},\\%\label{derivative_yy}\\
 \pmixdiff{A_{x}}{x}{y}&=4xy\pdoublediff{A_{x}}{\xi}+2x\pmixdiff{A_{x}}{\xi}{\eta}.\\%\label{derivative_xy}
\end{align*}
Substituting into Eq.~(\ref{big_pde}) and simplifying, we get
\begin{equation*}
 x^{2}\pdoublediff{A_{x}}{\eta}-y\pdiff{A_{x}}{\eta}+A_{x}=0
\end{equation*}
and completing the change of variables gives
\begin{equation} \label{1st_sub}
 \rbr{\xi-\eta^{2}}\pdoublediff{A_{x}}{\eta}-\eta\pdiff{A_{x}}{\eta}+A_{x}=0.
\end{equation}

The next step of the solution is to perform a reduction of order through the use of the known solution $A_{x}=\eta$ \cite{Kreyzig2006}. The reduction used is
\begin{equation*}
 A_{x}=u\eta,\quad\pdiff{A_{x}}{\eta}=\eta\pdiff{u}{\eta}+u,\quad\pdoublediff{A_{x}}{\eta}=\eta\pdoublediff{u}{\eta}+2\pdiff{u}{\eta},
\end{equation*}
which we substitute into Eq.~(\ref{1st_sub}) to give 
\begin{align*}\label{2nd_sub}
 \eta\rbr{\xi-\eta^{2}}\pdoublediff{u}{\eta}+2\rbr{\xi-\eta^{2}}\pdiff{u}{\eta}-\eta^{2}\pdiff{u}{\eta}-u\eta+u\eta&=0,\\
 \eta\rbr{\xi-\eta^{2}}\pdoublediff{u}{\eta}+\rbr{2\xi-3\eta^{2}}\pdiff{u}{\eta}&=0.
\end{align*}
We now make the substitution $v=\pdiff{u}{\eta}$,
\begin{equation} \label{3rd_sub}
 \eta\rbr{\xi-\eta^{2}}\pdiff{v}{\eta}+\rbr{2\xi-3\eta^{2}}v=0.
\end{equation}
This equation can now be solved separably,
\begin{equation*}
 \int\frac{1}{v}dv=\int\sbr{\frac{3\eta^{2}-2\xi}{\eta\rbr{\xi-\eta^{2}}}}d\eta.
\end{equation*}
We decompose the denominator through the use of partial fractions, giving
\begin{equation*}
 \int\frac{1}{v}dv=-\int\frac{2}{\eta}d\eta+\int\frac{\eta}{\xi-\eta^{2}}d\eta,\\
\end{equation*}
from which we get
\begin{align*}
 \ln{v}&=-2\ln{\eta}-\frac{1}{2}\ln{\abs{\xi-\eta^{2}}}+\ln{\sbr{\alpha\rbr{\xi,z}}},\\
 v&=\frac{\alpha\rbr{\xi,z}}{\eta^{2}\abs{\xi-\eta^{2}}^{\frac{1}{2}}},
\end{align*}
where $\alpha$ is an arbitrary function and we note that $\xi-\eta^{2}=x^{2}+y^{2}-y^{2}=x^{2}>0$, hence $\xi-\eta^{2}$ is
always positive.

Now that we have a solution to Eq.~(\ref{3rd_sub}), we must reverse our substitutions to get a solution for $A_{x}$. First, we integrate $v$ to get $u$,
\begin{equation}
 u=\alpha\rbr{\xi,z}\int\frac{1}{\eta^{2}\rbr{\xi-\eta^{2}}^{\frac{1}{2}}}d\eta.
\end{equation}
From standard integrals \cite{Gradshteyn2000}, we get
\begin{equation*}
 u=-\frac{\alpha\rbr{\xi,z}}{\xi\eta}\rbr{\xi-\eta^{2}}^{\frac{1}{2}}+\beta\rbr{\xi,z},
\end{equation*}
where $\beta$ is another arbitrary function. Next we write $A_{x}=u\eta$, obtaining
\begin{equation}
 A_{x}=\alpha\rbr{\xi,z}\rbr{\xi-\eta^{2}}^{\frac{1}{2}}+\eta\beta\rbr{\xi,z}
\end{equation}
where we absorb the factor of $-\frac{1}{\xi}$ into $\alpha$. Finally, we substitute back from $\rbr{\xi,\eta,z}$ to $\rbr{x,y,z}$,
\begin{align} \label{A_x_soln}
 A_{x}&=\alpha\rbr{x^{2}+y^{2},z}\rbr{x^{2}+y^{2}-y^{2}}^{\frac{1}{2}}+y\beta\rbr{x^{2}+y^{2},z},\nonumber\\
 &=x\alpha\rbr{x^{2}+y^{2},z}+y\beta\rbr{x^{2}+y^{2},z},
\end{align}
to give the solution for $A_{x}$. We find $A_{y}$ from Eq.~(\ref{sim_2}),
\begin{align*}
 A_{y}=&y\pdiff{A_{x}}{x}-x\pdiff{A_{x}}{y}\\
 =&y\sbr{\alpha\rbr{x^{2}+y^{2},z}+2x^{2}\alpha'\rbr{x^{2}+y^{2},z}+2xy\beta'\rbr{x^{2}+y^{2},z}}\\
 &-x\sbr{2xy\alpha'\rbr{x^{2}+y^{2},z}+\beta\rbr{x^{2}+y^{2},z}+2y^{2}\beta'\rbr{x^{2}+y^{2},z}}\\
 =&y\alpha\rbr{x^{2}+y^{2},z}-x\beta\rbr{x^{2}+y^{2},z},
\end{align*}
giving us solutions for all three components.

We will now verify that the solutions for $A_{x},A_{y},$ and $A_{z}$, satisfy the remaining equation~(\ref{sim_1}),
\begin{widetext}
\begin{align*}
 -x&\pdiff{}{y}\rbr{\pdiff{A_{x}}{x}+\pdiff{A_{y}}{y}+\pdiff{A_{z}}{z}}-x\pdiff{A^{2}}{y}+y\pdiff{}{x}\rbr{\pdiff{A_{x}}{x}+\pdiff{A_{y}}{y}+\pdiff{A_{z}}{z}}+y\pdiff{A^{2}}{x}\\
 =&-x\pdiff{}{y}\rbr{2\alpha+2x^{2}\alpha'+2xy\beta'+2y^{2}\alpha'-2xy\beta'+\pdiff{\gamma}{z}}+y\pdiff{}{x}\rbr{2\alpha+2x^{2}\alpha'+2xy\beta'+2y^{2}\alpha'-2xy\beta'+\pdiff{\gamma}{z}}\\
 &-x\rbr{4x^{2}y\alpha\alpha'+2y\beta^{2}+4y^{3}\beta\beta'+2y\alpha^{2}+4y^{3}\alpha\alpha'+4x^{2}y\beta\beta'+4y\gamma\gamma'}\\
 &+y\rbr{2x\alpha^{2}+4x^{3}\alpha\alpha'+4xy^{2}\beta\beta'+4xy^{2}\alpha\alpha'+2x\beta^{2}+4x^{3}\beta\beta'+4x\gamma\gamma'}\\
 =&-4xy\alpha'-4x^{3}y\alpha''-4xy\alpha'-4xy^{3}\alpha''-2xy\pdiff{\gamma'}{z}+4xy\alpha'+4xy\alpha'+4x^{3}y\alpha''+4xy^{3}\alpha''+2xy\pdiff{\gamma'}{z}=0,\\
\end{align*}
where a prime denotes differentiation with respect to $\rbr{x^2+y^2}$.

Therefore, the form of the vector potential required for $\sbr{\hat{H},\hat{L}_z}=0$ is,
\begin{equation} \label{nice_gauge}
 \mbf{A}=\sbr{x\alpha\rbr{x^{2}+y^{2},z}+y\beta\rbr{x^{2}+y^{2},z},y\alpha\rbr{x^{2}+y^{2},z}-x\beta\rbr{x^{2}+y^{2},z},\gamma\rbr{x^{2}+y^{2},z}}.
\end{equation}
This form of the vector potential satisfies the condition $\sbr{\hat{H},L_z}=0$ for wavefunctions with
$\pdiff{\psi}{x}, \pdiff{\psi}{y}$ and $\pdiff{\psi}{z}$ all independent of each other. Clearly, vector potentials that are not of
the form of Eq.~(\ref{nice_gauge}) do not satisfy the condition. However, in order to ensure that vector potentials
of this form satisfy this condition for an arbitrary wavefunction, we use the properties of the vector potentials to solve the original commutator
Eq.~(\ref{commute_eqn}) for an arbitrary wavefunction. This gives
\begin{align*} 
 \sbr{\hat{H},L_z}\psi=&2\rbr{x\alpha+y\beta}\pdiff{\psi}{y}-2\rbr{y\alpha-x\beta}\pdiff{\psi}{x}-2x\sbr{\rbr{2xy\alpha'+\beta+2y^{2}\beta'}\pdiff{\psi}{x}+\rbr{\alpha+2y^{2}\alpha'-2xy\beta'}\pdiff{\psi}{y}+2y\gamma'\pdiff{\psi}{z}}\\
 &+2y\sbr{\rbr{\alpha+2x^{2}\alpha'+2xy\beta'}\pdiff{\psi}{x}+\rbr{2xy\alpha'-\beta-2x^{2}\beta'}\pdiff{\psi}{y}+2x\gamma'\pdiff{\psi}{z}}\\
 &-x\psi\pdiff{}{y}\rbr{2\alpha+2x^{2}\alpha'+2xy\beta'+2y^{2}\alpha'-2xy\beta'+\pdiff{\gamma}{z}}\\
 &+y\psi\pdiff{}{x}\rbr{2\alpha+2x^{2}\alpha'+2xy\beta'+2y^{2}\alpha'-2xy\beta'+\pdiff{\gamma}{z}}\\
 &-x\psi\rbr{4x^{2}y\alpha\alpha'+2y\beta^{2}+4y^{3}\beta\beta'+2y\alpha^{2}+4y^{3}\alpha\alpha'+4x^{2}y\beta\beta'+4y\gamma\gamma'}\\
 &+y\psi\rbr{2x\alpha^{2}+4x^{3}\alpha\alpha'+4xy^{2}\beta\beta'+4xy^{2}\alpha\alpha'+2x\beta^{2}+4x^{3}\beta\beta'+4x\gamma\gamma'}.
\end{align*}
\end{widetext}
This simplifies to
\begin{align*}
 \sbr{\hat{H},L_z}\psi=&-x\psi\pdiff{}{y}\rbr{2\alpha+2x^{2}\alpha'+2y^{2}\alpha'+\pdiff{\gamma}{z}}\\
 &+y\psi\pdiff{}{x}\rbr{2\alpha+2x^{2}\alpha'+2y^{2}\alpha'+\pdiff{\gamma}{z}}\\
 =&-4xy\alpha'-4x^{3}y\alpha''-4xy\alpha'-4xy^{3}\alpha''-2xy\pdiff{\gamma'}{z}\\
 &+4xy\alpha'+4xy\alpha'+4x^{3}y\alpha''+4xy^{3}\alpha''+2xy\pdiff{\gamma'}{z}\\
 =&0.
\end{align*}
Thus, the vector potentials of the form~(\ref{nice_gauge}) fulfill the condition $\sbr{\hat{H},\hat{L}_z}=0$ and in these gauges $\hat{L}_{z}$ is a
constant of motion.

\subsection{Gauge transformations between vector potentials for which $L_{z}$ is a constant of motion}

We now consider a gauge transformation between two gauges of the form~(\ref{nice_gauge}).
A vector potential of this form gives the magnetic field
\begin{align*}
 \mbf{B}&=\nabla\times\mbf{A}\\
 &=\nabla\times\sbr{x\alpha+y\beta,y\alpha-x\beta,\gamma}\\
 &=\sbr{2y\gamma'-y\pdiff{\alpha}{z}+x\pdiff{\beta}{z},x\pdiff{\alpha}{z}+y\pdiff{\beta}{z}-2x\gamma',\right.\\
 &\qquad\left.2xy\alpha'-\beta-2x^{2}\beta'-2xy\alpha'-\beta-2y^{2}\beta'}\\
 &=\sbr{2y\gamma'-y\pdiff{\alpha}{z}+x\pdiff{\beta}{z},x\pdiff{\alpha}{z}+y\pdiff{\beta}{z}-2x\gamma',\right.\\
 &\qquad\left.-2\beta-\rbr{2x^{2}+2y^{2}}\beta'}.
\end{align*}
Since any modification to $\beta$ would affect the $-2\beta$ term in the $z$ component of $\mbf{B}$ and $\mbf{B}$ must be unchanged by
gauge transformations, $\beta$ must be constant in a gauge transformation.

A gauge transformation is given by $\mbf{A'}=\mbf{A}+\nabla\chi$ and takes the form

\begin{align} \label{delta_A}
 \nabla\chi&=\mbf{A'}-\mbf{A},\nonumber\\
 &=\sbr{x\Delta\alpha+y\Delta\beta,y\Delta\alpha-x\Delta\beta,\Delta\gamma}\nonumber\\
 &=\sbr{x\Delta\alpha,y\Delta\alpha,\Delta\gamma},
\end{align}
using that $\Delta\beta$ must be zero. We obtain $\chi$ by integrating each of the components of the vector~(\ref{delta_A}):
\begin{align*}
 \int x\Delta\alpha dx=\frac{\lambda\rbr{x^{2}+y^{2},z}}{2},\\
 \int y\Delta\alpha dy=\frac{\mu\rbr{x^{2}+y^{2},z}}{2},\\
 \int\Delta\gamma dz=\nu\rbr{x^{2}+y^{2},z}.
\end{align*}
Clearly then, the scalar field $\chi$ must be a function of the form $\chi\rbr{x^{2}+y^{2},z}$.

Finally, we demonstrate that $\sbr{\mbf{r}\times\mbf{j}_{p}}_{z}$ is unchanged by gauge transformations between gauges of the form~(\ref{nice_gauge}).
The paramagnetic current density transforms according to $\mbf{j}_{p}'=\mbf{j}_{p}+\rho\nabla\chi$,
\begin{align*}
 \sbr{\mbf{r}\times\mbf{j}_{p}'}_{z}&=\sbr{\mbf{r}\times\rbr{\mbf{j}_{p}+\rho\nabla\chi}}_{z}\\
 &=\sbr{\mbf{r}\times\mbf{j}_{p}}_{z}+\sbr{\mbf{r}\times\rho\nabla\chi}_{z}\\
 &=\sbr{\mbf{r}\times\mbf{j}_{p}}_{z}+\sbr{\mbf{r}\times\rho\sbr{x\Delta\alpha,y\Delta\alpha,\Delta\gamma}}_{z}\\
 &=\sbr{\mbf{r}\times\mbf{j}_{p}}_{z}+\rho\sbr{xy\Delta\alpha-xy\Delta\alpha}\\
 &=\sbr{\mbf{r}\times\mbf{j}_{p}}_{z}.
\end{align*}
So, when we are in any gauge of the form of Eq.~(\ref{nice_gauge}), and when we transform between any of these gauges, both $\hat{L}_{z}$ and
$\sbr{\mbf{r}\times\mbf{j}_{p}}_{z}$ are unaffected.

\bibliography{References}
\bibliographystyle{unsrt.bst}

%\begin{acknowledgments}
% Put your acknowledgments here.
%\end{acknowledgments}

\end{document}